\documentclass[draft]{agujournal2019}
\usepackage{url} %
\usepackage{lineno}
\usepackage{soul}
\usepackage{placeins}

\draftfalse

\journalname{JGR: Solid Earth}

\newcommand{\degree}{$^\circ$}

\begin{document}

\title{Characterising the Atacama segment of the Chile subduction margin (24\degree S - 31\degree S) with $>$ 165,000 earthquakes}

\authors{
Jannes Münchmeyer\affil{1},
Diego Molina-Ormazabal\affil{1},
David Marsan\affil{1},
Mickaël Langlais\affil{1},
Juan-Carlos Baez\affil{2},
Ben Heit\affil{3},
Diego González-Vidal\affil{4},
Marcos Moreno\affil{5},
Frederik Tilmann\affil{3,6},
Dietrich Lange\affil{7},
Anne Socquet\affil{1}
}

\affiliation{1}{Univ. Grenoble Alpes, Univ. Savoie Mont Blanc, CNRS, IRD, Univ. Gustave Eiffel, ISTerre, Grenoble, France}
\affiliation{2}{National Seismological Center, University of Chile, Santiago, Chile}
\affiliation{3}{GFZ, German Research Centre for Geosciences, Telegrafenberg, 14473, Potsdam, Germany}
\affiliation{4}{Departamento Ciencias de la Tierra, Facultad de Ciencias Químicas, Universidad de Concepción}
\affiliation{5}{Department of Structural and Geotechnical Engineering, Pontificia Universidad Católica de Chile, Santiago, Chile}
\affiliation{6}{Institute for Geological Sciences, Freie Universität Berlin, Berlin, Germany}
\affiliation{7}{GEOMAR Helmholtz Centre for Ocean Research Kiel, Wischhofstr. 1-3, 24148 Kiel, Germany}

\correspondingauthor{Jannes Münchmeyer}{munchmej@univ-grenoble-alpes.fr}

\begin{keypoints}
\item We created a detailed seismicity catalog for the Atacama gap with more than 165,000 events using almost 200 seismic stations.
\item The dense seismicity describes the geometry and structure of the subduction from outer rise to deep earthquake nests and upper plate faults.
\item The very low location uncertainties ($<$50~m) show a fine-scale, fractal segmentation of the interface into seismic and aseismic regions.
\end{keypoints}

\newcommand{\citeneeded}{\textsuperscript{[\textcolor{blue}{citation needed}]} }

\begin{abstract}  %
The Atacama segment in Northern Chile (24\degree S to 31\degree S) is a mature seismic gap with no major event ($M_w \geq 8$) since 1922.
In addition to regular seismicity, around the subducting Copiapó ridge, the region hosts seismic swarms, and shallow and deep slow slip events.
To characterize the fine structure of this seismic gap and its seismic-aseismic interplay, we instrumented the region with almost 200 seismic and geodetic stations.
Using machine learning, we derived a dense, high-resolution seismicity catalog, encompassing over 165,000 events with double-difference relocated hypocenters.
Our catalog details the outer rise, interface, intraslab, crustal and mantle wedge seismicity.
We infer a detailed slab geometry, showing that the flat slab is dipping towards the south with a narrower extent along dip.
The slab geometry controls the intraslab seismicity, with cross-cutting activity in the region of highest bending and a downdip limit around 105~km slab depth.
Our catalogue exhibits significant seismicity in the mantle wedge upper corner between 28\degree S and 31\degree S, highlighting the brittle behavior of the cold nose.
On the subduction interface, interplate locking controls the updip end of the seismicity, with seismicity extending closer to the trench in low-locking areas.
On fine scales, resolved by relative uncertainties below 50~m, the subduction interface has a complex 3D structure, showing a fractal distribution of seismic patches down to a scale of tens of meters.
Our results provide a holistic view of this complex subduction zone, while at the same time giving insights into fine-scale structures and processes.
\end{abstract}

\section*{Plain Language Summary}

In Northern Chile, the Nazca plate subducts underneath the South American plate, regularly producing major earthquakes.
However, some regions have not experienced a large earthquake in over a century.
One such region is the Atacama segment in Northern Chile, with the last big event in 1922.
To better understand this region, we deployed several temporary seismic networks to complement the permanent networks, totaling almost 200 stations.
In this paper, we use their waveforms to create a dense seismic catalog with more than 165,000 events.
We use recently developed machine learning techniques to obtain a very complete catalog with precise event locations.
The seismicity in this region consists of several distinct classes: the outer rise seismicity west of the trench, the interface seismicity at shallow to intermediate depth, the intraslab seismicity in the subducting plate, and the upper plate seismicity.
In addition, we catalog more than 12,000 blasting events from open pit mining.
Our catalog helps understanding the structure of the subduction in the Atacama segment, such as the geometry of the subducting slab and the segmentation of the plate interface.
We make our catalog openly available, to enable future, more detailed studies.

\section{Introduction}

\begin{figure*}[ht!]
\centering
\includegraphics[width=\textwidth]{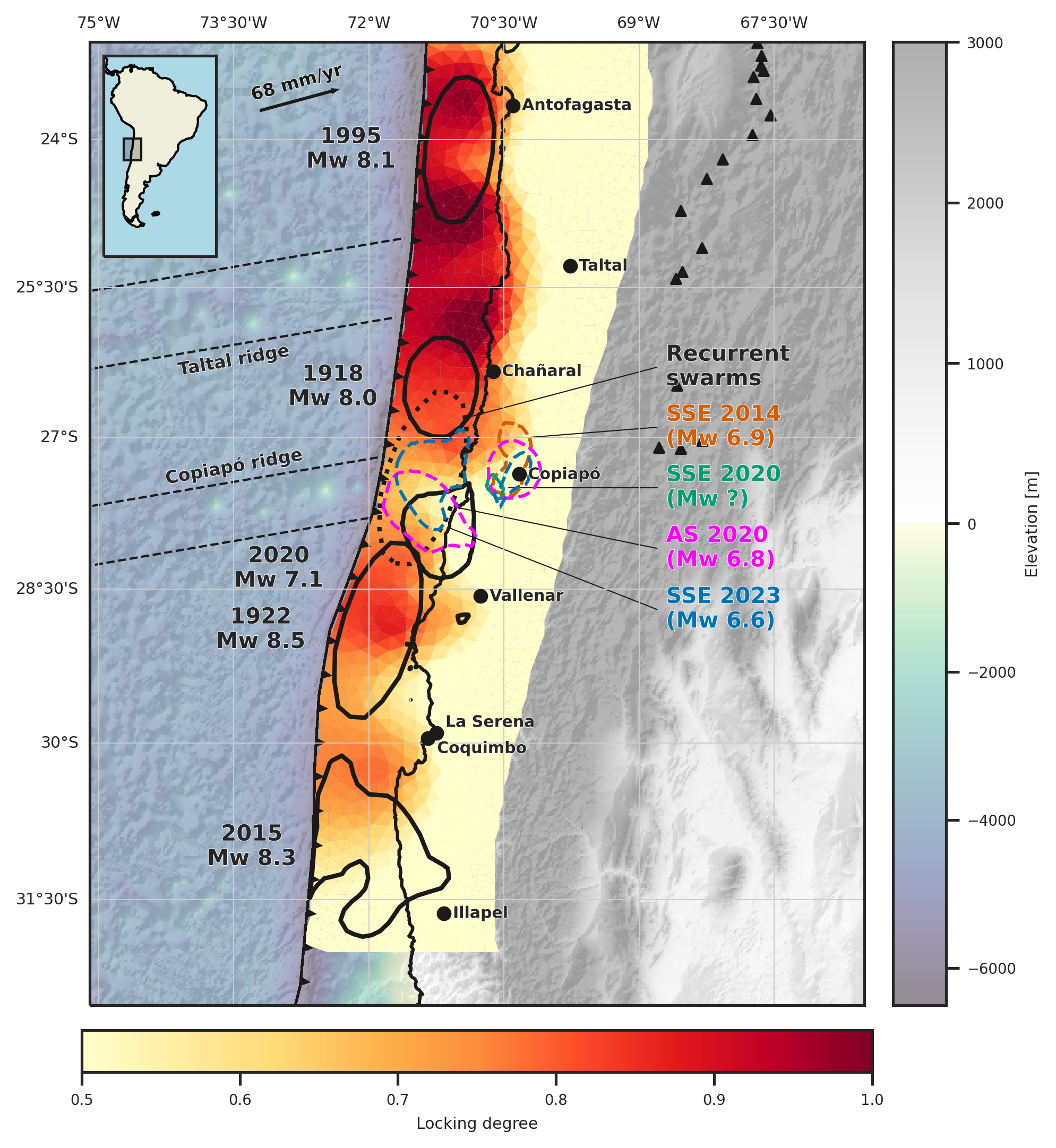}
\caption{Overview of the study region and surroundings. Yellow to orange shading shows interplate locking \cite{yanez-cuadraInterplateCouplingSeismic2022}. Major earthquakes rupture areas are outlined in solid black \cite{ruizHistoricalRecentLarge2018,kleinComprehensiveAnalysisIllapel2017,vignySearchLostTruth2024}. Recurrent swarm activity is shown by a dotted black outline \cite{comteSeismicityStressDistribution2002,holtkampEarthquakeSwarmsSouth2011,ojedaSeismicAseismicSlip2023,marsanEarthquakeSwarmsChilean2023,munchmeyer2024chile_sse}. Slow slip events (SSEs) are shown by dashed colored outlines \cite{kleinDeepTransientSlow2018,kleinReturnAtacamaDeep2022,munchmeyer2024chile_sse}. An additional dashed line shows the afterslip (AS) of the 2020 Huasco sequence \cite{molina2024sse}. Active volcanoes are shown as black triangles. Major cities are shown as black dots with labels. 
For a map depicting the seismic networks and study region boundaries, see Figure~\ref{fig:station_map}.}
\label{fig:seismotectonic_map}
\end{figure*}

The subduction margin in Northern Chile is one of the world's most active seismic regions (Figure~\ref{fig:seismotectonic_map}).
Here, the Nazca plate subducts beneath the South American plate with a velocity of 68~mm/year \cite{norabuenaSpaceGeodeticObservations1998}.
Within the last 30 years, the region hosted three events with $M_w \geq 8$: the 1995 Antofagasta earthquake ($M_w=8.0$, $\sim23.5$\degree S), the 2014 Iquique earthquake ($M_w=8.2$, $\sim19.5$\degree S), and the 2015 Illapel earthquake ($M_w=8.3$, $\sim31.5$\degree S).
The northernmost part of the margin (18\degree S to 24\degree S) has been densely instrumented with seismic networks since 2007, yielding detailed and long-duration seismic catalogs for the region and illuminating the structure of the subduction \cite{sipplSeismicityStructureNorthern2018,sipplNorthernChileForearc2023}.
However, the segment between 24\degree S and 31\degree S, the so-called Atacama segment, is substantially less well-described.

The Atacama segment, between 24.5\degree S and 30\degree S, has historically hosted major earthquakes, yet no earthquake with $M_w \geq 8$ has occurred since 1922 \cite{comteSeismicityStressDistribution2002,ruizHistoricalRecentLarge2018,vignySearchLostTruth2024}.
Therefore, this region is also known as the 1922 or Atacama gap.
The gap is delineated to the north by the 1995 Antofagasta earthquake and to the south by the 2015 Illapel earthquake.
At the latitude of Copiapó, the segment hosts recurring deep and shallow slow slip events (SSEs) (Figure~\ref{fig:seismotectonic_map}, colored contours) \cite{kleinDeepTransientSlow2018,kleinReturnAtacamaDeep2022,munchmeyer2024chile_sse}.
The shallow SSEs are accompanied by intense seismic swarm activity \cite{comteSeismicityStressDistribution2002,ojedaSeismicAseismicSlip2023,munchmeyer2024chile_sse}.
To constrain the fine structure of this seismic gap and understand this mature seismic gap and the complex interactions of slow and fast deformation, several initiatives installed seismo-geodetic deployments starting in late 2020 \cite{fdsn_9c,fdsn_xz,fdsn_y6,fdsn_3v}.
\citeA{gonzalez-vidalRelationOceanicPlate2023} used an early subset of this data to infer a catalog with $>$30,000 events spanning 20 months.
Yet, a longer duration, more detailed catalog is still missing.

While the slab in the northernmost part of Chile (18\degree S - 24\degree S) shows little heterogeneity along strike, the Atacama segment is characteristed by a more complex slab geometry.
First, the direction of the trench changes around 28\degree S.
Second, south of 27\degree S the region is characterized by a section of flattening slab \cite{pardoSeismotectonicStressDistribution2002,hayesSlab2ComprehensiveSubduction2018a}.
Furthermore, two oceanic ridges enter the subduction (Figure~\ref{fig:seismotectonic_map}): the Taltal ridge (around 25.5\degree S) and the Copiapó ridge (around 27.5\degree S).
The age of the subducting Nazca plate at the trench in this region has been estimated at $\sim 43$~Ma, making this region an intermediate subduction zone in terms of thermal structure \cite{vankekenSubductionFactoryDepthdependent2011}.
Due to the arid climate, the subduction is sediment-starved, with a sediment coverage at the trench of only $\sim200$~m \cite{vankekenSubductionFactoryDepthdependent2011}.

Interseismic locking around the 1922 gap shows strong along-strike variation \cite{yanez-cuadraInterplateCouplingSeismic2022,gonzalez-vidalRelationOceanicPlate2023} (Figure~\ref{fig:seismotectonic_map}).
The margin is highly locked (locking degree $>0.8$) north of 27.5\degree S with the highest locking around 26\degree S.
An area of low locking in front of the Copiapó ridge (27.5\degree S to 28\degree S, locking down to 0.6) separates this patch from another highly locked patch ($>0.8$).
Locking drops substantially south of 29\degree S with values consistently below 0.7.
Comparing to the seismicity, \citeA{gonzalez-vidalRelationOceanicPlate2023} find that the low-locking channel corresponds to the highest rate of shallow interface seismicity, while the downdip edge of the highly locked patches corresponds to an increase in intermediate depth seismicity.

The region of low locking between 27.5\degree S and 28\degree S coincides with the subducting Copiapó ridge.
It has been identified as a seismic barrier for several megathrust events \cite{comteSeismicityStressDistribution2002,ruizHistoricalRecentLarge2018}.
As it is debated, whether the 1922 earthquake broke through this area \cite{vignySearchLostTruth2024}, it is unclear how persistent this barrier is.
At the same time, this region hosts offshore seismic swarms, with records dating back as far as 1973 \cite{comteSeismicityStressDistribution2002,holtkampEarthquakeSwarmsSouth2011,ojedaSeismicAseismicSlip2023,marsanEarthquakeSwarmsChilean2023}.
Recent results suggest that these are related to aseismic deformation from shallow slow slip events \cite{munchmeyer2024chile_sse} and extensive afterslip \cite{molina2024sse}.
Even further downdip, deep slow slip events with a recurrence interval of approximately 5 years and a duration of over a year have been observed in 2006, 2010, 2014 and 2020 \cite{kleinDeepTransientSlow2018,kleinReturnAtacamaDeep2022,molina2024sse}.
The joint occurrence of these diverse types of slow and fast earthquakes renders the area particularly interesting for investigating the underlying mechanisms.

In this study, we derive a high-density seismicity catalog combing data from several temporary and permanent deployment.
Our catalog contains more than 165,000 double-difference relocated hypocenters, covering the time between November 2020 and February 2024.
We describe the catalog in detail, study the seismicity patterns, and infer structural properties of the subduction zone and, in particular, the plate interface.
Furthermore, we derive a new, refined slab surface model from the catalog.
Our results provide insights into the complex subduction environment of the Atacama seismic gap and the interaction between seismic and aseismic processes.
At the same time, we hope that our catalog can serve as a basis for future studies of the subduction in the Atacama segment.

\section{Data and Methods}

\subsection{Data}

\begin{figure*}[ht!]
\centering
\includegraphics[width=\textwidth]{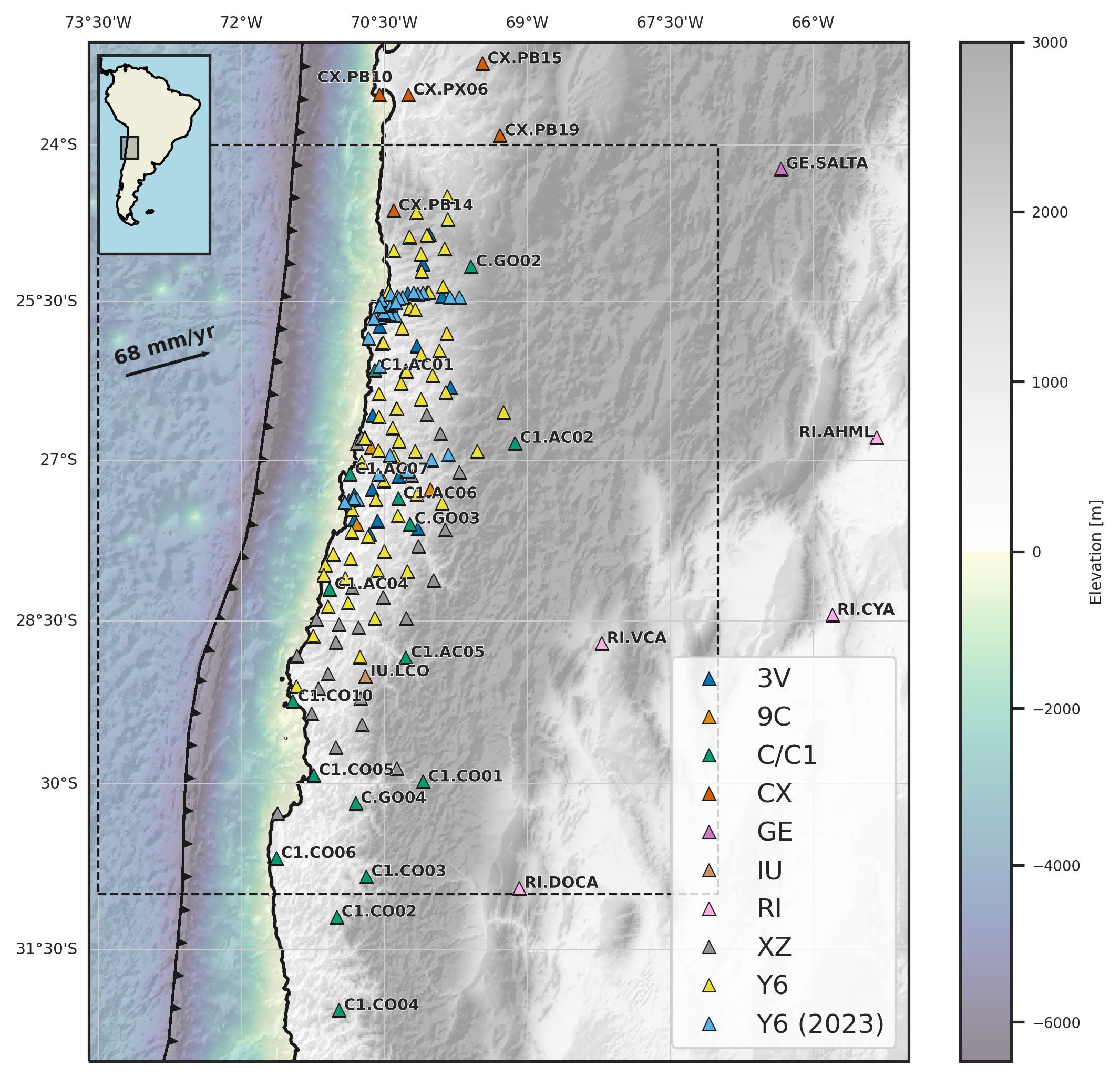}
\caption{Overview map with the stations incorporated into the catalog generation. Stations are colored according to their seismic network. For stations from permanent networks, the station codes are provided. Within the Y6 network, stations deployed for 4 months in 2023 are indicated separately. Additional IPOC stations (network CX) to the north of the mapped region have been used for the catalog. The black, dashed rectangle marks the study area.}
\label{fig:station_map}
\end{figure*}

The earthquake catalog spans 40 months (November 2020 to February 2024) and is based on temporary and permanent deployments.
As the deployment periods differ between the networks, the station coverage is highly variable throughout the duration of the catalog.
In this section, we provide an overview of the incorporated data, while in the results section, we highlight the resulting spatial distribution of magnitude of completeness over time.
A map of all 193 stations is provided in Figure~\ref{fig:station_map}.
Data availability for all stations is visualised in Figure~\ref{fig:data_availability}.

As a backbone, we use permanent stations from the C (4 stations), C1 (17) \cite{fdsn_c1}, CX (23) \cite{fdsn_cx}, IU (1) \cite{fdsn_iu}, GE (1) \cite{fdsn_ge} and RI (4) networks.
Excluding occasional downtimes, these stations ran for the whole study duration providing a basic coverage.
A substantially denser coverage is provided by a collection of temporarily installed networks.
The ANILLO deployment (Y6) \cite{fdsn_y6} lasted from November 2020 to April 2022 and is itself composed of two phases.
In the first phase (November 2020 to March 2021) 10 broadband sensors and 19 short-period sensors have been deployed.
In the second phase (March 2021 to April 2022), the network was extended to a total of 20 broadband sensors, 19 short-period sensors, and 20 geophones.
The DEEPtrigger network (XZ) \cite{fdsn_xz} was deployed between June and August 2021 and operated continuously until February 2024.
It consisted of 17 Trillium Compact 20s Posthole broadband sensors and 8 CMG40 short-period instruments.
The IMO network (3V) \cite{fdsn_3v} was deployed in December 2022 and provided data until November 2023.
The network encompassed 26 PE-6 4.5Hz geophones.
In the period from February 2023 to June 2023, 30 geophones have been deployed to register shots from a seismic experiment offshore \cite{lange2023high}.
No data from the ocean bottom seismometers associated with this cruise has been incorporated in this study.
From the 9C network \cite{fdsn_9c}, we use 3 broadband stations in the Copiapó area.
We include 9C data from November 2020 to November 2023 in this study.

\subsection{Event detection, location and characterisation}

We use a multi-step approach for building the earthquake catalog (Figure~\ref{fig:workflow}).
From the continuous waveforms, we detect P and S arrivals using the deep learning picker PhaseNet \cite{zhuPhaseNetDeepneuralnetworkbasedSeismic2019}.
We use the PhaseNet model integrated in SeisBench \cite{woollamSeisBenchToolboxMachine2022} and trained on the INSTANCE dataset from Italy \cite{micheliniINSTANCEItalianSeismic2021}.
This model has shown high performance even when applied to other regions and tectonic settings \cite{munchmeyerWhichPickerFits2022}.
We use a picking threshold of 0.1 for both P and S phases and a temporal overlap of 50~\% between subsequent input windows of PhaseNet.

We associate the picks with the PyOcto associator \cite{munchmeyerPyOctoHighthroughputSeismic2024} using a 1D velocity model and station travel time residuals.
For association, we required at least 10 picks per event and 4 stations with both P and S pick.
While we enforce more strict quality control criteria for the final catalog, we retain the larger number of events from association throughout the processing to aid relative methods.

We perform two location steps.
First, we calculate absolute locations for each event individually using NonLinLoc \cite{lomaxProbabilisticEarthquakeLocation2000}.
Second, we use GrowClust3D \cite{trugmanGrowClust3DJlJulia2022} to perform relative relocation of all events jointly.
For GrowClust3D, we combine differential travel times from picked arrivals with cross-correlation times.
For each event, we calculate differential times to its 500 nearest neighbors (up to a distance of 50~km) using cross-correlation.
We correlate P arrivals on the vertical component and S arrivals on the horizontal components.
We select windows comprised of 0.5~s before and 1.0~s after the picked arrival bandpass-filtered between 1 and 10~Hz.
We use the correlation routine in obspy, a time-domain correlation that is refined to subsample precision through parabolic interpolation, allowing a maximum shift of 0.3~s.
We only keep event pairs with a correlation value above 0.7.
In total, we use 14.2 million differential travel times from cross-correlation.
On top of these, we add differential times from picked phase arrivals such that each event has at least 25 neighbors with 10 differential travel times.
This is primarily necessary for events far outside the network.

We calibrate a local magnitude scale using a simplified version of the method from \citeA{munchmeyerLowUncertaintyMultifeature2020}.
For each S pick, we infer the peak displacement on the combined horizontal components after simulating the response of a Wood-Anderson seismometer.
Based on the recorded amplitudes, we infer a 1D attenuation curve with respect to the hypocentral distance and station residuals.
We jointly optimize for the attenuation curve and the station corrections using minibatch gradient descent.
We chose the constant offset term of the attenuation curve in a way that minimises bias to the magnitudes reported in the CSN catalog.
The attenuation curve is visualised in Figure~\ref{fig:attenuation_curve}.
In this study, we use the mean magnitude across all stations.
In addition, we provide median values and station magnitude estimates in the published catalog.
Observed magnitude uncertainties, inferred from the standard errors across the recording stations, are generally below 0.1 magnitude units throughout the study area.
The reported magnitude values will saturate for large events for two reasons.
First, the frequency band of the Wood-Anderson instrument does not cover the lower frequencies of large events.
Second, we do not incorporate strong motion data, i.e., might use a small number of clipped records.
Within the scope of this study, these limitations are not problematic as we focus on structural aspects of the seismicity distribution.

We identify events related to mine blasting using a heuristic approach.
We split all events shallower than 30~km into 0.2$^\circ$ by 0.2$^\circ$ grid cells.
If within the grid cell, the rate of activity during the daytime substantially exceeds (5x) that rate at night, all events in the cell are labeled as mine blasts.
To avoid edge effects, we perform the same analysis with a grid shifted by 0.1$^\circ$ in each direction.
We label events as potential mine blasts if they have been identified as blasting in at least one of the two grids.
For validation, we compare the identified mining sites to satellite imagery.
In all checked cases, open-pit mines are present.
In addition to mine-blasts, we detected airgun shots from an active seismic cruise in March 2023 \cite{lange2023high}.
We identified and labeled these manually based on the cruise report.

For the final catalog, we keep all events with at least 12 picked phases and at least 5 stations where both a P and an S wave has been picked.
Furthermore, we require standard errors in the NonLinLoc origin time below 0.5~s.
We keep events irrespective whether they could be relocated with GrowClust3D or not to avoid biasing the catalog against isolated events.

To obtain adequate velocity models and station corrections, we execute the workflow three times.
We perform a first run up to NonLinLoc using the 1D velocity model from \citeA{graeber1999three}.
Based on this first iteration of the catalog, we infer new velocity models.
We first infer a new 1D model using Velest \cite{kisslingInitialReferenceModels1994}.
Based on this model, we infer a 3D velocity model using simul2017 \cite{eberhart-phillipsSimul2017FlexibleProgram2021}.
We use a 120~km node spacing in North-South direction and a 25~km node spacing in East-West direction to account for the higher variability in velocity along dip than along strike.
Along the vertical axis, we use an adaptive spacing ranging from 5~km at shallow depth ($<20$~km) to 40~km below 200~km depth.
For both Velest and simul2017, we only use events with high numbers of picks (40/30 picks, 10/8 stations with P and S picks for Velest/simul2017) and low location residuals (0.4/0.5 RMSE NonLinLoc).
We perform 100/128 runs with random subsamples containing 2500 events to infer mean models and standard deviations.
For simul2017, we use a stratified sampling procedure to ensure consistent ray coverage.
To this end, we split the region into disjoint 1\degree $\times$ 1\degree ~cells and draw an equal number of events from each cell.
The resulting models are visualised in Figures~\ref{fig:velocity_vp},~\ref{fig:velocity_dvp},~\ref{fig:velocity_vpvs} and are available in the supplementary material.
We then perform a second run of the phase association (using the new 1D model) and location with NonLinLoc (using the new 3D model) to infer station travel-time residuals.
With these residuals and the velocity models, we perform a final complete run of all processing steps using the 3D model for NonLinLoc and GrowClust3D.

We used an HPC cluster with servers with 16 CPU cores for catalog generation.
Phase picking and amplitude extraction took 6 days of server time, phase association and location with NonLinLoc one day each, cross-correlation 13 days, and all remaining tasks together less than one day.
We parallelised the computation across 45 servers using dask \cite{dask2016} to achieve a total wall time below 24 hours.

For an analysis on the long-term stability of the seismicity distribution, we create an additional reference catalog.
The reference catalogs spans the time from 2014 to February 2024, but only includes data from permanent stations.
We use the same processing pipeline for the reference catalog as for the main catalog.
For locating events, we use the 3D velocity model and station residuals inferred using the temporary networks.
Relative relocation is performed independently for the reference catalog and the main catalog.

\section{Catalog characteristics and quality}

\subsection{Location uncertainties}

Figure~\ref{fig:nlloc_uncertainties} visualises the absolute location uncertainties estimated with NonLinLoc.
For shallow events ($<$ 50~km) near or within the network, median uncertainties vary between 1 and 4~km (vertical and horizontal).
Uncertainties for events in the outer rise are around 4~km horizontally and 10~km vertically. 
Within the network, uncertainties stay below 3~km (vertical and horizontal) even for events up to 100~km depth.
Outside the network, in the deep clusters to the east, horizontal uncertainties are between 10 and 25~km, while vertical uncertainties can be even higher.
This is expected, as the network does not provide sufficient coverage in this area.
We provide a detailed comparison of our location estimates with the catalog of the Chilean Seismic Network (CSN) and the Global CMT catalogs in supplementary text S1.

We estimated relative errors through multiple analysis.
First, we visually inspected the sharpness of features.
For interface events within the network, we are able to distinguish substructures within the interface at a scale of tens of meters (Section~\ref{sec:interface}).
We complement this analysis with a study of fractal dimensions (Section~\ref{sec:segmentation}), suggesting that errors within the network are below 50~m.
For a more quantitative measure, we perform a bootstrap analysis using GrowClust3D.
The details of the bootstrap are described in \cite{trugmanGrowClustHierarchicalClustering2017}.
As performing a bootstrap analysis for the whole catalog at once is computationally infeasible, we subselect a roughly 1 by 1 degree area around 27.5\degree S.
This area hosts a dense offshore seismic swarm on the plate interface, a challenging yet interesting case for relocation.
As GrowClust uses a hierarchical clustering approach to relocation, our analysis of a subregion should not yield different results than bootstrapping on the whole catalog.

Figure~\ref{fig:relocation_bootstrap} shows the results of the bootstrapping analysis.
Comparing different iterations of the bootstrapping, we see that while the individual runs show shifts in absolute location, their relative locations are very stable.
As expected relative positions between events show higher variability between events located further apart than for closely colocated events.
To quantify this relationship, for each event pair we compare the media interevent distance across the bootstrap iterations to the median absolute deviation (MAD) of the distance, a measure of the uncertainty (Figure~\ref{fig:relocation_bootstrap}c).
We use hypocentral distances, i.e., taking both epicenter and depth into account.
At an interevent distance of 500~m, the median MAD is below 20 meter, and the $90^{th}$ percentile at 50~m.
Even at 5~km interevent distance, the $90^{th}$ percentile MAD only exceeds 100~m slightly, even though we observe individual outliers from poorly linked events.

These low errors might seem surprising, especially for a cluster outside the network.
However, they are the effect of multiple factor.
First, the quality of relative relocations depends crucially on the number of events because each additional event provides a reference frame for the others.
For $n$ events, the degrees of freedom for location is on the order of $4n$ (latitude, longitude, depth, time), while the number of event pairs, i.e., potentially independent travel time measurements from cross-correlation, grows with $n^2$.
As machine learning catalogs, like the one presented in this study, enable a very high event density, they are expected to have low relative location errors.
To profit even more from this effect, we perform the relocation on a larger catalog ($\sim330,000$~events) before subselecting events for the final catalog.
Second, our relocation uses differential times from both P and S phases.
This provides distance constraints, that are particularly useful outside the network.
Third, the data quality is high, as the Atacama segment has low background noise and the geology leads to few near-site effects.
This allows to measure precise differential travel times using cross-correlations.
Lastly, in our relocation we use a 3D velocity model inferred from the data.
This leads to accurate estimates of the slowness values around the events, lowering systematic errors.
We emphasis that all of these factor decrease the relative location error, but only the 3D velocity model will decrease the absolute error.
Therefore, the relative location error will be substantially below the absolute errors.

\subsection{Magnitude of completeness}

\begin{figure*}[ht!]
\centering
\includegraphics[width=1\textwidth]{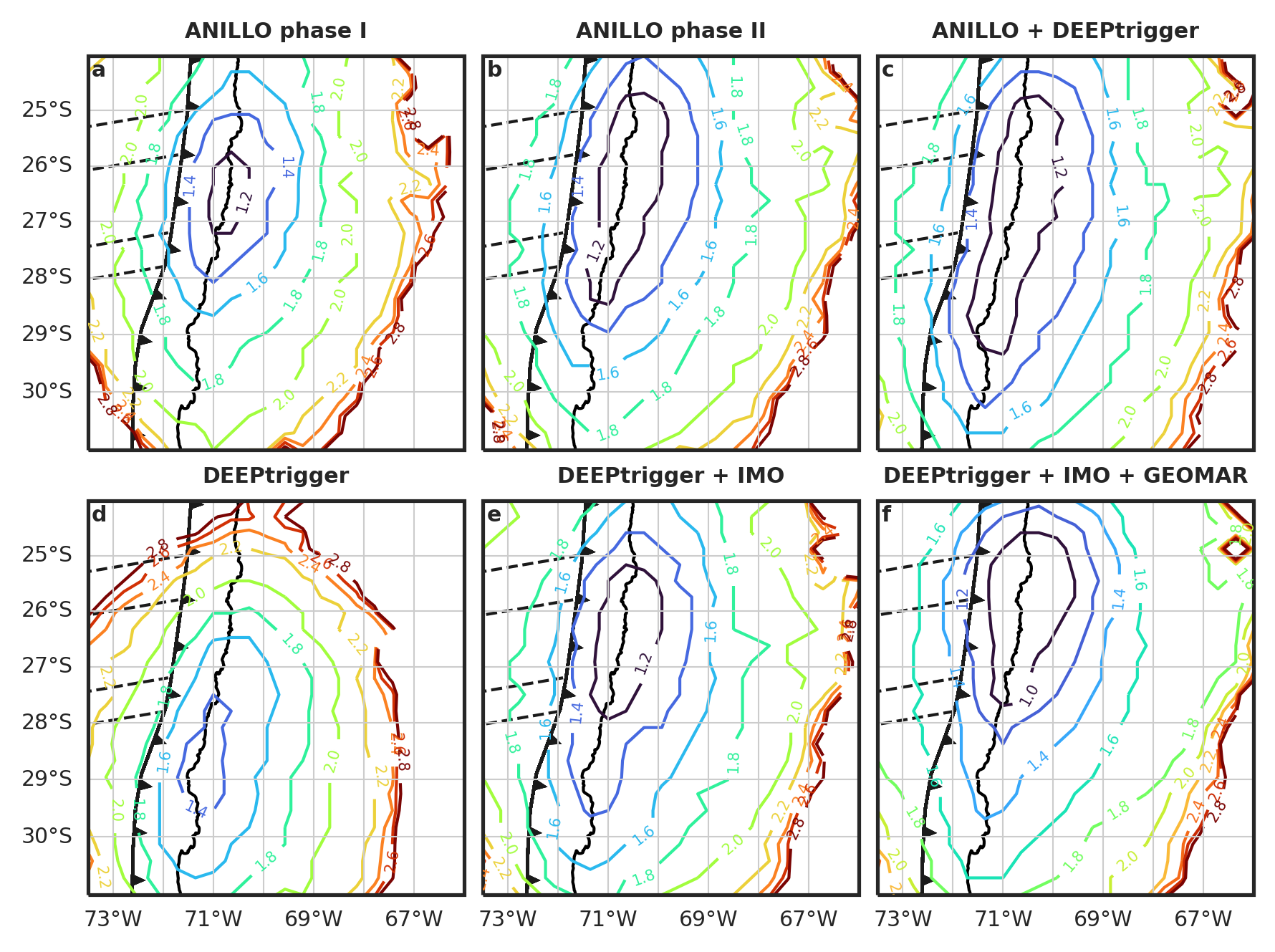}
\caption{Magnitude of completeness during the different stages of the deployment. For each location, the magnitude of completeness has been estimated at the 90$^{th}$ percentile of the depth of the events around this location. For deeper events, magnitude of completeness will be higher, for shallower events lower. Note that the maps show typical completeness during the deployments but that completeness is varying on a daily basis due to transient station unavailability.}
\label{fig:completeness_maps}
\end{figure*}

Figure~\ref{fig:magnitude_histogram} shows the magnitude distribution split by event category.
The overwhelming majority of the events falls between magnitudes 1 and 6.
The magnitude of completeness $M_c$ varies with the extent of the seismic networks in different deployment periods and spatially due to the large size of the study area.
To estimate $M_c$ and its spatial variation for the different deployment periods, we use a modified version of the method by \citeA{schorlemmerProbabilityDetectingEarthquake2008}.
Details about the estimation of $M_c$ are provided in supplementary text S2.

Figure~\ref{fig:completeness_maps} shows the spatial distribution of $M_c$.
Due to the variable station coverage over time, $M_c$ shows temporal variations.
Generally, $M_c$ is highest close to the coastline.
During the most dense deployment, $M_c$ between 25\degree S and 27\degree S goes down to 1.0.
The ANILLO + DEEPtrigger and DEEPtrigger + IMO deployments allow a coverage of 1.6 along almost the whole coastline within a  band of 3 degree width.
$M_c$ for outer rise events varies between 1.6 and 2.0.
With increasing distance to the main network, $M_c$ reduces downdip to between 1.6 and 2.0.

\subsection{Event classification}

To aid a detailed analysis, we infer a slab interface model from the seismicity.
We manually pick the slab surface between 24\degree S and 31\degree S from the trench to 67\degree W on a grid with 20 x 50 grid points.
We only pick the slab interface where sufficient seismicity is available.
The resulting slab interface is interpolated and smoothed.
For picking in the shallow section, we identify the slab surface through seismicity clusters that are oriented parallel to the expected interface, allowing to separate upper plate events from interface seismicity.
This strategy is necessary as we do not have focal mechanism information for our catalog.
After the downdip end of the interface seismicity, we pick the slab top at the upper end of the intraslab seismicity.
This leads to an apparent step in the slab model that we smooth out.
Excerpts from the resulting slab interface model are visualised in Figure~\ref{fig:overview} and will be discussed later.
The slab interface model is available in the supplementary material.

Based on this slab geometry, we classify the events into six classes.
We use the following criteria, based on the structures observed in the catalog.
The criteria are applied in order, i.e., events are assigned to the first class matching the criteria.
\begin{itemize}
 \item Events between 2023-03-07 and 2023-03-21, shallower than 30~km, between 25.15\degree S and 25.55\degree S, and between 70.8\degree W and 70.1\degree W are labeled airgun shots (517 events).
 \item Events identified as potential mine blasts are labeled mine blast (12,296 events).
 \item Events west of the trench are labeled outer rise (3,290 events).
 \item Events at least 3~km above the slab surface are labeled upper plate (18,537 events).
 \item Events between 3~km above and 4~km below the slab surface and at locations where the slab surface is at most 60~km deep are labeled interface (36,165 events).
 \item All remaining events, i.e., events below the slab surface or events near the slab surface at slab depths above 60~km are labeled intraslab (93,732 events).
\end{itemize}

In addition, we label some events towards the east outside the network as mislocated.
We assign this label for events east of 69\degree W that are deeper than 25~km and at least 10~km above the slab surface.
The locations for these events have been constrained without phase picks from the RI stations located above the seismicity.
Therefore, they have very poor depth constrains.
In total, this group contains 1,866 events.
We exclude these events from our subsequent analysis.

We note that this classification is only an approximation to aid the subsequent analysis.
In particular, the split into upper plate, interface, and intraslab events can be difficult in places with low vertical separation.
This is particularly evident for events located close to the trench, where the depth uncertainties are highest.
Similarly, the lower bound on interface events at 60~km depth is an approximation with the actual boundary varying along dip.
Nonetheless, this classification, with the criteria fixed based on visual inspection, is useful for describing the seismicity patterns.

\section{Structure of the subduction}

\subsection{Seismicity overview}

\begin{figure*}[ht!]
\centering
\includegraphics[width=1\textwidth]{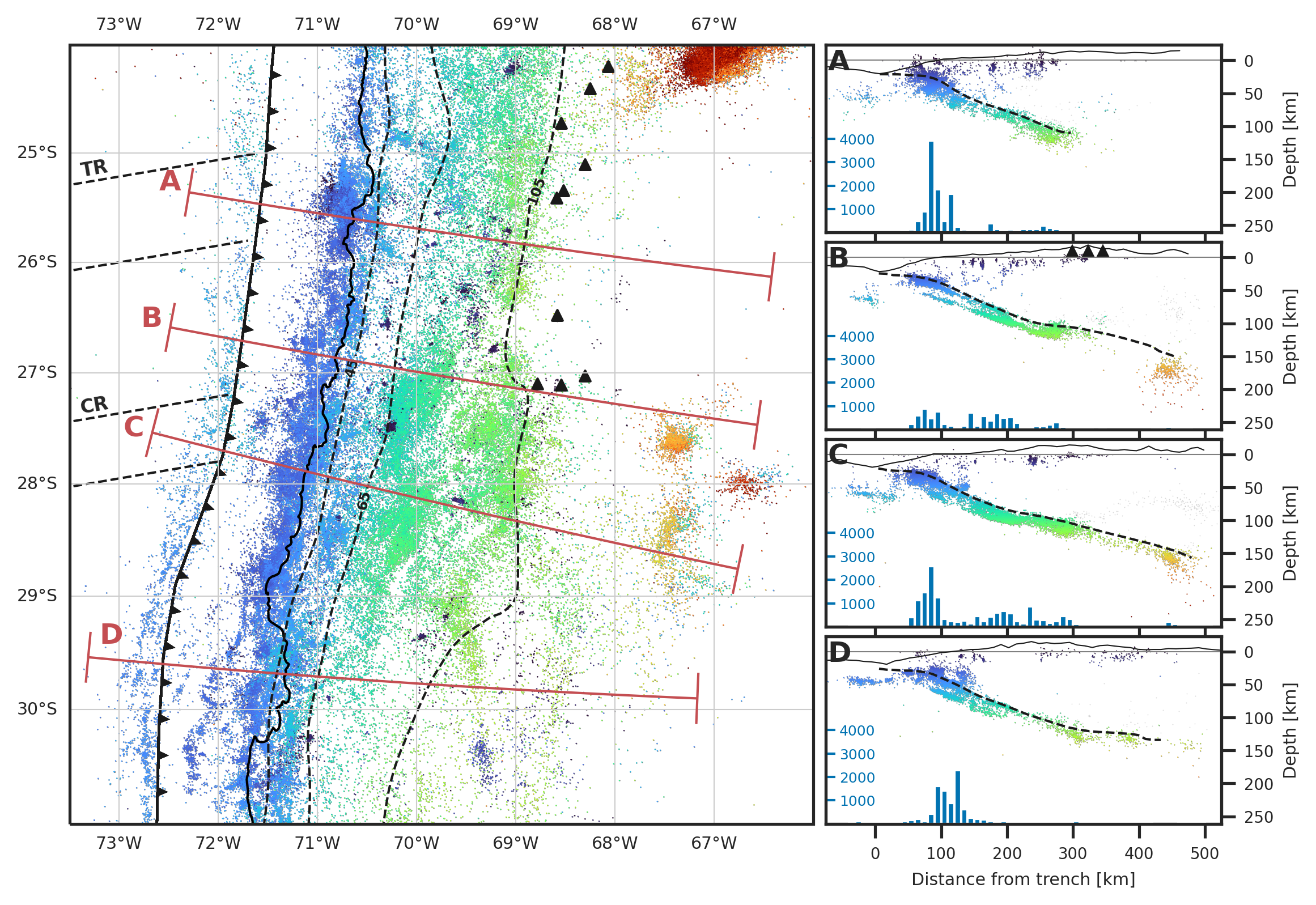}
\caption{Overview of the seismicity in the study area. Each dot represents an earthquake, colored by event depth. The red lines indicated the cross sections on the right. Each cross-section is centered along the line and has a width of 50~km, indicated by the whiskers. Cross sections are orthogonal to the trench. In the map, we show 45, 65 and 105~km iso-depth lines of the slab model inferred from the seismicity. Volcanoes are shown with black triangles. In the cross-sections, black dashed lines indicate the inferred slab model. The topography/bathymetry is three times exaggerated. The number of events per 10~km bin along dip are indicated with histograms in the cross-sections. We provide additional trench-perpendicular cross-sections in Figures~\ref{fig:events_overview_fine} and \ref{fig:cross_sections_fine}}
\label{fig:overview}
\end{figure*}

In total, our catalog comprises 166,403 events within the study region with $\sim$6.7~million associated phase picks (3.9~million P phases, 2.8~million S phases).
On average, each event is constrained by 40.1 phase picks.
Figure~\ref{fig:overview} shows a map view and selected cross-sections of the seismicity.
We first provide an overview of the event distribution before studying the seismicity in detail in the subsequent sections.
The events can be grouped into different categories according to the map view.
West of the trench, the catalog contains outer rise seismicity.
Underneath the coastline, a dominant band of shallow seismicity on the interface (up to 45~km depth) is detected.
We refer to this seismicity as shallow in contrast to the intermediate depth seismicity, while acknowledging that this set will include many events on the deeper part of the seismogenic plate interface.
Almost no seismicity is observed between the outer rise and the shallow seismicity, except in the region of the subducting Copiapó ridge (around 27.5\degree S) and around 30\degree S.
In these regions, moderate levels of seismicity exist throughout between outer rise and shallow seismicity.

Below the interface seismicity, intraslab seismicity in the oceanic plate makes up the majority of events in our catalog.
In most regions, the intraslab seismicity consists of two slab-parallel bands that merge below an interface depth of about 65~km.
They stretch out downdip to a slab depth of 130~km.
The along-strike distribution of intermediate depth seismicity is heterogeneous with the highest activity rates between 27\degree S and 29\degree S.
Intermediate seismicity levels are especially low south of 29\degree S.
Further downdip, below the 130~km mark, seismicity occurs within four distinct clusters, located around 24\degree S (the Jujuy cluster) and between 27.5\degree S and 28.5\degree S (the Pipanaco nest consisting of 3 clusters).
In addition to the lower plate and interface events, we observe seismic activity in the upper plate, combining crustal seismicity and mantle wedge seismicity.

The seismicity structure in the Atacama segment is similar to the seismicity in the adjacent segments to the north \cite{sipplNorthernChileForearc2023} and to the south \cite{sipplMicroseismicityAppearsOutline2021}.
Both areas are characterised by a layering of seismicity around the slab.
While the layering into three layers (interface and two intraslab) is well-resolved to the north of our region, the resolution to the south, in the metropolitan segment, is insufficient to separate the interface and the upper layer clearly.
The ratio between the amount of interface seismicity and seismicity within the slab varies along strike.
To the north, intraslab seismicity at intermediate depth generally makes up the majority of seismic events.
Further south, interface seismicity makes up a larger chunk of the total seismicity.
Upper plate seismicity occurs inland along large parts of the margin.
In addition, in the north, a dense seismic cluster of seismicity has been reported in the crustal wedge of the upper plate \cite{sipplSeismicityStructureNorthern2018}, terminating at 21.5\degree S.
As the cluster coincides with an area of low plate locking, it has been suggested that the seismicity is driven by ascending fluids \cite{blochHighresolutionImageNorth2014}.
While the long-term catalog of the CSN \cite{barrientosSeismicNetworkChile2018} reports some outer rise seismicity, none of the more comprehensive catalogs include it, presumably because the locations far offshore did not meet the quality control criteria of these catalogs.

We contextualise our results with the focal mechanisms from the Global Centroid Moment Tensor (GCMT) project (Figures~\ref{fig:events_overview_mt},~\ref{fig:cross_section_ortho_mt}) \cite{ekstromGlobalCMTProject2012}.
Up to 45~km interface depth, we observe predominantly thrust mechanisms compatible with the interface seismicity.
The orientation of the moment tensors becomes less uniform with higher vertical separation to the interface, however, due to the high depth uncertainties in GCMT it remains unclear in which structures these events locate.
Further downdip, at slab depths that do not support interface seismicity, we observe a change in focal mechanisms with predominantly normal faulting.
These downdip events also show stronger strike-slip components than their shallow counterparts.
As an exception from this observation, the steepest slab segment (between 25.5\degree S and 26.5\degree S) shows predominantly reverse mechanisms, suggesting a systematic difference in stressing due to the slab geometry compared to the surrounding segments.
The deep nests are dominated by normal mechanisms.
These mechanisms contain a stronger strike-slip component in the Pipanaco nest than in the Jujuy nest.

\subsection{Slab geometry}

\begin{figure*}[ht!]
\centering
\includegraphics[width=0.6\textwidth]{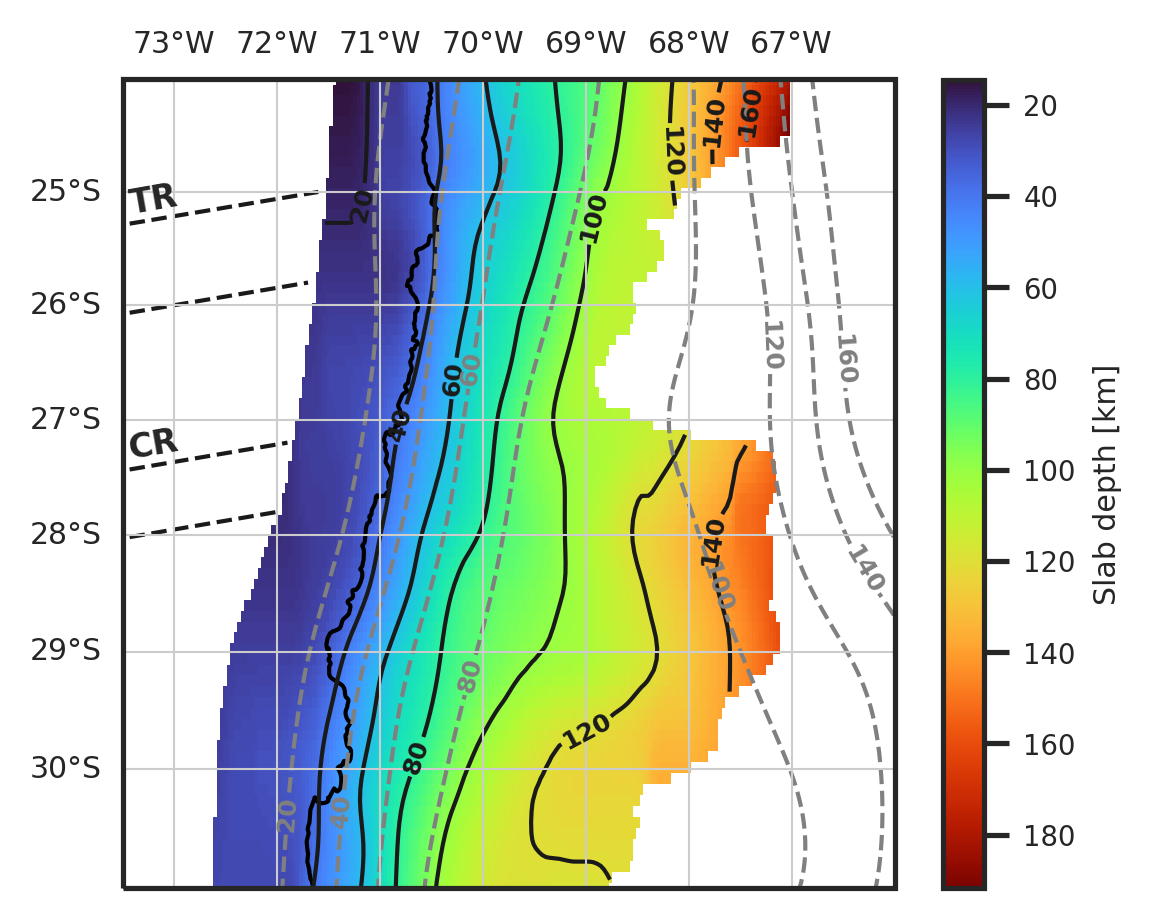}
\caption{Slab interface depth inferred from the seismicity. Coloured shading and black isolines indicate the inferred slab model. Grey dashed lines show the slab depth according to the slab2 model \cite{hayesSlab2ComprehensiveSubduction2018a}. No slab depth is indicated west of the trench and towards the east in regions with insufficient seismicity.}
\label{fig:slab_models}
\end{figure*}

We compare our slab model derived from the seismicity to slab2 \cite{hayesSlab2ComprehensiveSubduction2018a}, a global model based on a metastudy of seismicity and imaging techniques (Figure~\ref{fig:slab_models}).
Overall, our model reports a deeper slab, reaching from few kilometers at shallow depth to more than 20~km difference at 120~km depth.
This is similar to the previous study of \citeA{gonzalez-vidalRelationOceanicPlate2023}, who also find a deeper slab than slab2.
Around 27\degree S, the slab transitions from steeply dipping in the north towards a flatter slab in the south.
The flat section covers a depth range of 90 to 120~km and extends about 200~km along dip and at least 300~km along strike.
The flat portion occurs about 20~km deeper south of 29\degree S than to the North.
Further east, a clear dipping of the slab is indicated by the Pipanaco nest around 67\degree W between 28\degree S and 29\degree S.

We note that the absolute depth of our slab model is not reliable close to the trench.
While the depth here should approach the depth of the trench, around 7~km, this can not be reconciled with the observed seismicity pattern.
This is caused by the large vertical uncertainties outside the network, leading to a consistent overestimation of the depth (Figure~\ref{fig:nlloc_uncertainties}).
We decided to report the slab consistent to the seismicity to aid classification, instead of adjusting it to the bathymetry.
As seismic profile for the Copiapó region has recently been acquired \cite{warwel2024structure}, but is not included in this study and also not in a larger scale metastudy on the slab geometry \cite{contreras-reyesOffshoreGeometrySouth2025}.

While the overall trends are similar, our model shows notable differences in slab geometry compared to slab2.
Differences occur, in particular, for the flat slab south of 27\degree S.
First, it is less wide along dip than previously imaged and does not show a fully flat section at all.
While slab2 reports a steepening only east of 67\degree W, we observe the steepening already at 68\degree W.
Second, we image substantial along-strike variation in the depth of the flat section with a deepening towards the south.
In contrast, slab2 reports a consistent plateau between 80 and 100~km depth with its largest along-dip extent around 30.5\degree S.
Due to lacking seismicity and as it is outside the region with robust depth determinations, we are not able to identify the full extent of this deeper part of the flat slab to the south and further downdip.

Our model overlaps not only with slab2, but also with the models from \citeA{mulcahyCentralAndeanMantle2014} to the north and \citeA{bianchiTeleseismicTomographySouthern2013} and \citeA{andersonGeometryBrittleDeformation2007} to the south.
Of these, the first two only map the deeper part of the slab ($>$100~km).
The model from \citeA{mulcahyCentralAndeanMantle2014}, inferred from earthquake hypocenters, agrees with the steeper descent of our slab compared to slab2.
However, it still maps a wide extent for the flat slab along the dip direction.
To the south, the model from \citeA{bianchiTeleseismicTomographySouthern2013}, inferred from teleseismic receiver functions, confirms the smaller extent of the flat slab found in this study in the north-south direction.
It also agrees with the overall deeper slab and the deepening of the flat slab towards the south.
Similarly, the model by \citeA{andersonGeometryBrittleDeformation2007}, while only overlapping with our model by about 1 degree, shows a narrow flat slab around 30\degree S with a depth around 110 to 120~km.
This is consistent with our results.

While our seismicity catalog is highly detailed, we note shortcomings of our slab model.
First, the spatial coverage is limited by the availability of seismicity.
In particular in the east, this does not allow to infer slab geometries in all places.
Second, depth accuracy degrades in regions with poor azimuthal coverage.
While in the east, the stations from the Argentinian network provide good constraints, towards the trench, an accurate estimate of slab depth is not possible.
This would require active or passive seismic data from marine stations.
Third, the seismicity varies along dip.
While between 20 and 45~km depth, the interface shows dense seismic activity, allowing to accurately pick the slab top, further downdip the interface seismicity disappears.
This leads to an apparent step in slab depth at the downdip end of the interface seismicity.
At deeper depths, only intraslab seismicity occurs.
As it is unclear what the distance between the top of the intraslab seismicity and the slab top is, we pick the deeper part of the slab at the top of the intraslab seismicity.
We smoothly interpolate between these regimes to obtain a consistent geometry.

\subsection{Seismicity within the subducting plate}

\begin{figure*}[ht!]
\centering
\includegraphics[width=\textwidth]{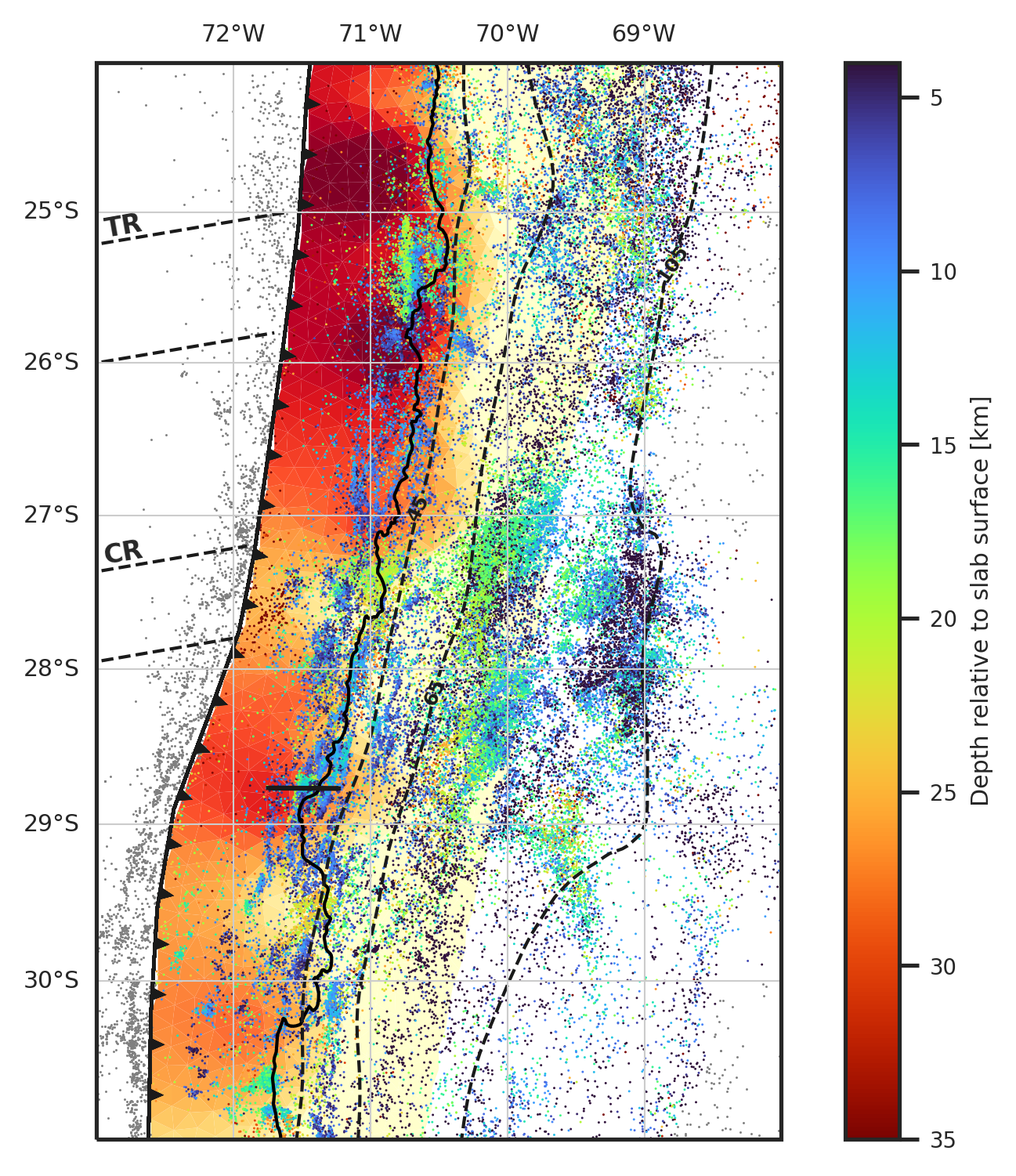}
\caption{Seismicity within the subducting slab (at least 4~km below slab top). The color indicates the relative depth to the slab surface. Outer rise events are indicated in grey for orientation. Black lines indicate the trench and depth contours of the slab. A black line denotes the profile in Figure~\ref{fig:cross_section_example_combined}. The profile width is not indicated, as it is smaller than the line width.}
\label{fig:slab_seismicity}
\end{figure*}

\begin{figure*}[ht!]
\centering
\includegraphics[width=\textwidth]{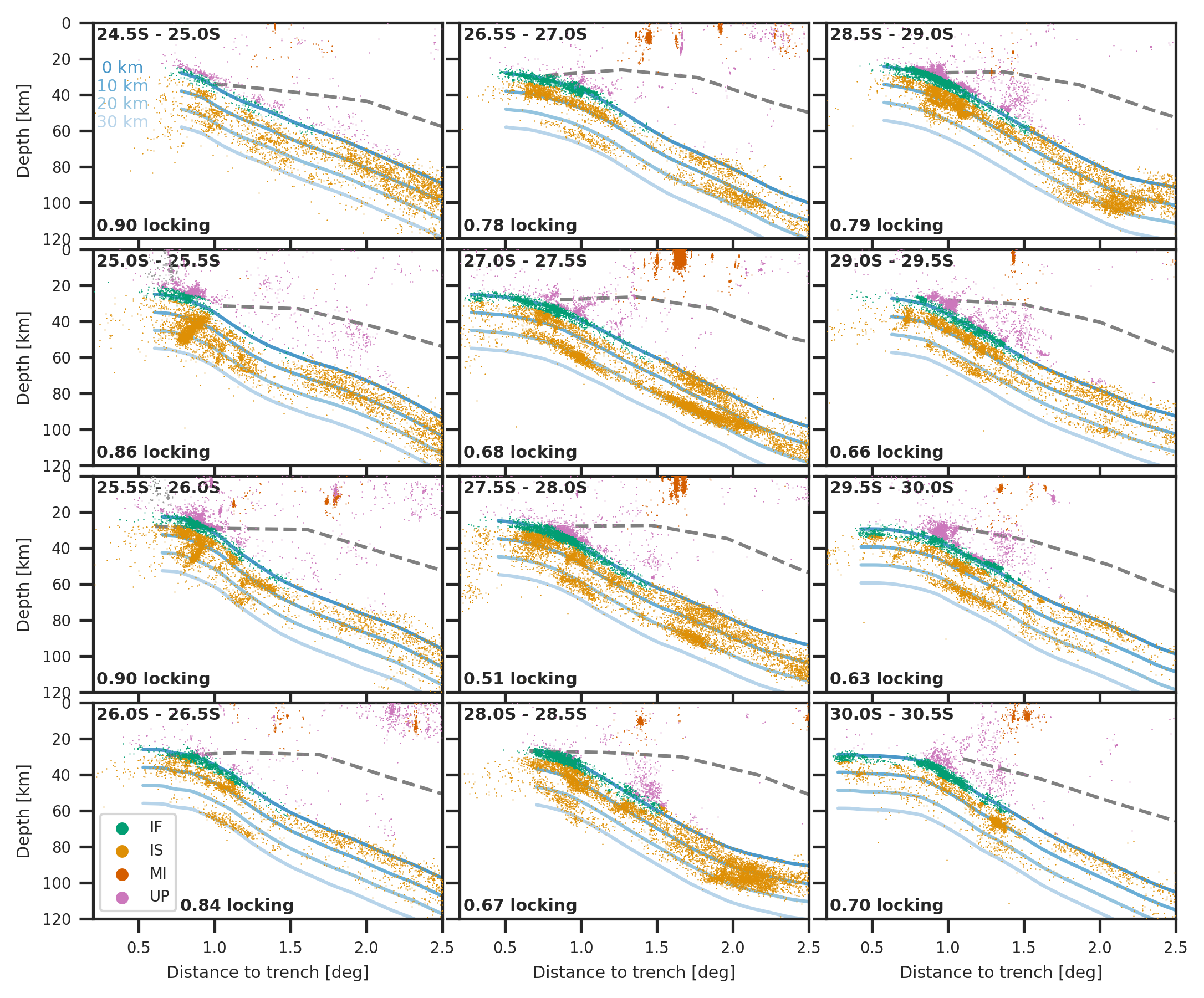}
\caption{Cross-sections of the seismicity along latitude. Events are colored according to their class (IF - interface, IS - intraslab, MI - mining, UP - upper plate). Cross-sections are zoomed in to focus on the shallower seismicity, excluding outer rise events and deep clusters. Blue lines visualise the slab interface model inferred from the data and parallels with 10~km, 20~km, and 30~km vertical offset. We only visualise the slab interface model where it is constrained by the data. Grey dashed lines indicate the continental Moho based on \citeA{rivadeneyra-veraUpdatedCrustalThickness2019}. Average plate locking according to \citeA{yanez-cuadraInterplateCouplingSeismic2022} within 1.5 degrees of the trench is provided in the lower left.}
\label{fig:cross_section_ortho}
\end{figure*}

About 60~\% of the seismicity in our catalog occurs within the subducting slab ($\sim$94,000~events), so called intraslab seismicity.
Figure~\ref{fig:slab_seismicity} shows these intraslab events color-coded by their vertical distance to the slab surface, while Figure~\ref{fig:cross_section_ortho} (orange dots) shows cross-sections.
In shallow to intermediate depth regions (slab surface down to 65~km), we observe intraslab seismicity with a clear vertical separation (5 to 10~km) to the interface seismicity.
These events show two different patterns north and south of 26\degree S.
In the north, we observe a dense, cross-cutting cluster of seismicity oriented perpendicular to the slab, reaching from roughly 5~km to 30~km below the slab (Figure 7, 25\degree S to 26\degree S).
This feature coincides with the strongest bending of the slab in the model we inferred, suggesting a relation between the slab bending and the cross-cutting seismicity \cite{raneroRelationshipBendfaultingTrenches2005}.

In contrast, further south the seismicity aligns in two slab-parallel layers, each with a vertical extent of up to $\sim$5~km (Figure 7, e.g., 27.0\degree S to 27.5\degree S).
The separation of these layers to the interface and to each other varies, with the center of the upper layer locating between 5 and 10~km below the interface and the lower layer between 20 and 25~km.
While we do not have a detailed thermal model available for this region, the comparison to a model for the adjacent segment in the north \cite{cabreraNorthernChileIntermediatedepth2021} suggests that the lower layer of the seismicity coincides with the 600\degree C to 650\degree C isotherms, i.e., the temperature range for the antigorite transition.
Studies of comparable double seismic zones suggest that the two planes experience extensional and compressional stresses from plate bending and are separated by a neutral stress plane \cite{kitaExistenceInterplaneEarthquakes2010,sipplGlobalConstraintsIntermediateDepth2022}.
The activity rate appears correlated between the two layers but not between the intraslab and interface seismicity.
The intraslab seismicity is mostly diffuse and does not show clearly visible fine-scale structures.
This reflects in the activity rates along strike showing smooth variation.
Nonetheless, some structures are visible in the intraslab seismicity distribution such as the strike-parallel lineaments at 28\degree S-29\degree S, 70.5\degree S-71\degree W.

Downdip, intraslab activity rates are low between 45~km and 65~km slab top depth (Figure~\ref{fig:slab_seismicity}).
The two layers of seismicity dissolve around 65~km slab top depth, except for the latitude ranges between 26.5\degree S and 28.0\degree S and 29.0\degree S and 29.5\degree S, where the layering only dissolves at slab depth below 80~km (Figure~\ref{fig:cross_section_ortho}).
Activity rates in this depth range vary along strike, with highest seismic activity between 26.5\degree S and 29.0\degree S.
While there is further activity to the North, only smaller amounts of intraslab activity occur south of 29.0\degree S.
The highest activity rates coincides with the flat section of the slab (26.5\degree S and 29.0\degree S).
At the downdip edge, the activity is largely constrained by the 105~km isodepth line of the slab surface.

At slab depth below 105~km, we observe seismicity exclusively in four distinct seismicity clusters (Figure~\ref{fig:overview} red and orange dots).
One cluster is located at the edge of our study region, around 24\degree S 67\degree W, and is known as the Jujuy seismic nest \cite{cahillSeismicityShapeSubducted1992,mulcahyCentralAndeanMantle2014,valenzuela-malebranSourceMechanismsRupture2022}.
It locates at a depth between 150~km and 250~km and is highly seismically active, with more than 100 M4+ EQs a year.
\citeA{valenzuela-malebranSourceMechanismsRupture2022} suggest that the seismicity in this nest is driven by dehydration of lithospheric serpentinite processes and subsequent fluid ascent.
Due to its location outside the network, we are unable to constrain potential internal structures.

The three other clusters locate in close proximity to each other downdip the flat slab region between 27.5\degree S and 28.5\degree S.
These clusters are known as the Pipanaco nest \cite{mulcahyCentralAndeanMantle2014}.
Each of the clusters is roughly circular with a radius between 20 and 40~km.
Two clusters occur around 150~km depth with an along-strike separation of about 100~km.
Another 60~km downdip and between the two other clusters along strike, a third, deeper cluster occurs.
While these three clusters stand out from the seismically quiet surrounding areas, they are by far not as active as the Jujuy nest.

Comparing the activity along the subducting plate to neighboring segments of the megathrust, we observe clear similarities to the section to the north.
\citeA{sipplNorthernChileForearc2023} report a similar segmentation of the intraslab seismicity into two distinct layers between 18\degree S and 24\degree S.
However, further north this separation is more stable than in our study area, with a typical separation of 7~km between interface and upper layer and 25~km between interface and lower layer.
At greater depth, the merging of the two layers happens almost simultaneously everywhere along strike, in contrast to the along-strike variation in our region.
This difference is likely caused by the different slab geometry in the Atacama segment.
While further north the subduction has almost constant strike and little geometrical variation along strike, in the Atacama segment we observe higher complexity with a varying strike and the flat slab segment.

\subsection{Continental crust and mantle wedge seismicity}

\begin{figure*}[ht!]
\centering
\includegraphics[width=\textwidth]{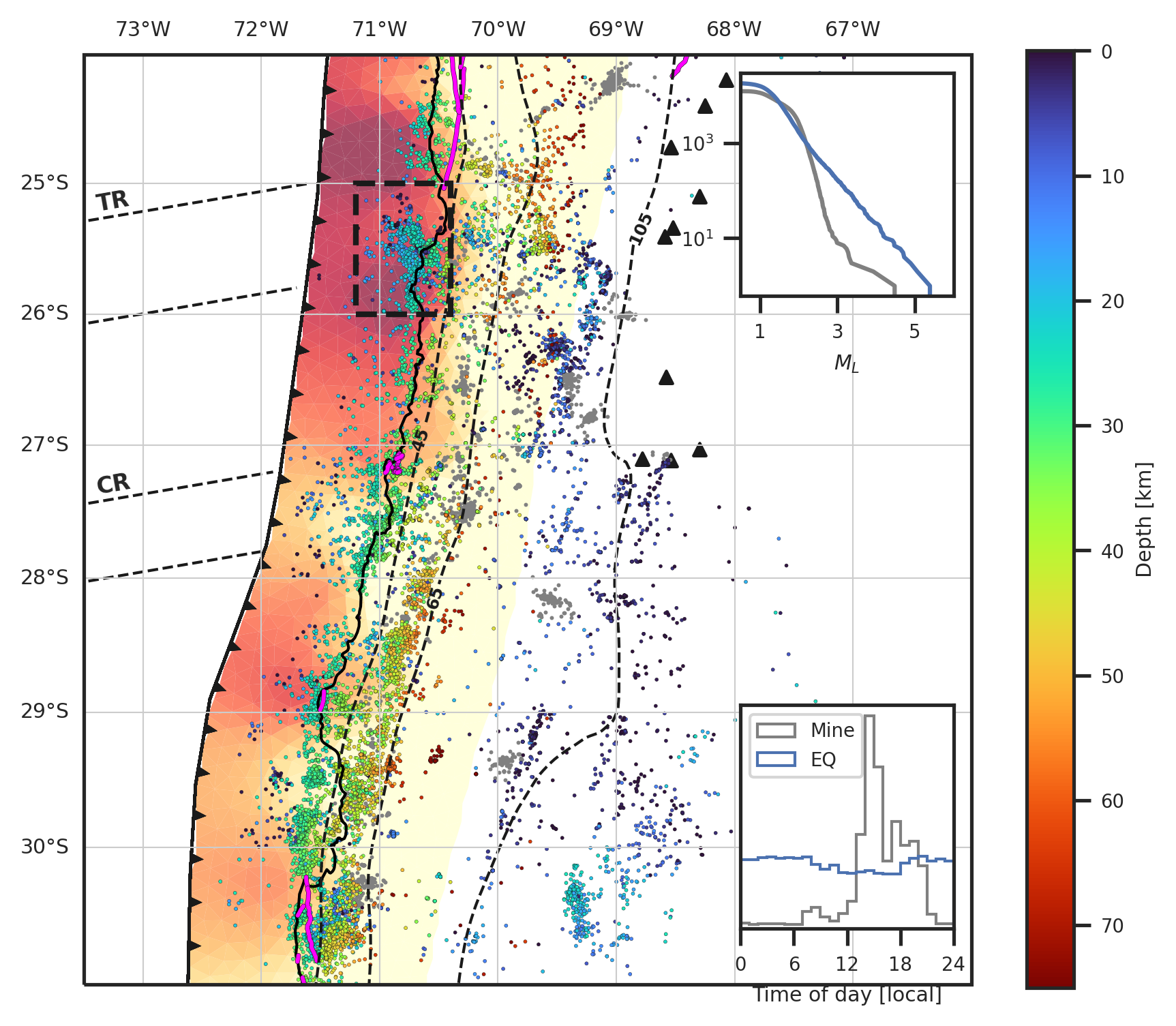}
\caption{Seismicity in the upper plate (more than 3~km above the slab and at most 75~km deep). Earthquakes are shown in color, mine blasts in grey, active crustal faults from \citeA{maldonadoComprehensiveDatabaseActive2021} in magenta. Interseismic locking from \citeA{yanez-cuadraInterplateCouplingSeismic2022} is shown in the background. The bottom inset shows histograms for the local time of day at which events in both classes occur. The top inset shows the magnitude distribution for both classes. Black lines indicate the trench and depth contours of the slab. The continental Moho intersects the slab around the 45~km depth contour. Black triangles denote active volcanoes. Cross-sections for the area marked by a black dashed rectangle are available in Figure~\ref{fig:cross_section_crustal}}
\label{fig:upper_plate}
\end{figure*}

In total, we detect around 30,800 events in the upper plate, of which 12,300 have been identified as mine blasts, leaving 18,500 earthquakes (pink in Figure~\ref{fig:cross_section_ortho}, Figure~\ref{fig:upper_plate}).
The insets in Figure~\ref{fig:upper_plate} validate our separation of mining and tectonic sources, even though individual events might still be misclassified.
We observe different classes of upper plate seismicity.
First, we see shallow crustal seismicity onshore (Figure~\ref{fig:upper_plate}, blue colors).
North of 27\degree S, shallow crustal faulting occurs in the western cordillera, especially between 2\degree ~and 3\degree ~from the trench, terminating just west of the volcanic arc.
This is consistent with topography building through transfer of the constrains from the subduction interface to the overriding plate \cite{armijoCoupledTectonicEvolution2015}.
South of 27\degree S, where the volcanic arc disappears because of the flat slab, the shallow crustal seismicity is distributed over a wider area extending up to 4\degree ~from the trench.
This seismicity aligns along a series of en echelon NNE-SSW structures, consistent with the predominant direction of the topography (Figure~\ref{fig:upper_plate_topography}, e.g., 27\degree S-28\degree S, 68\degree W-70\degree W).
This suggests right lateral movement in the cordillera, potentially associated to a slightly oblique convergence rate in this area generating slip partitioning between the subduction and the upper plate structures.
We find little overlap with mapped surface faults (Figure~\ref{fig:upper_plate}, magenta lines), yet this is likely caused by the low density of mapped faults in the region \cite{mulcahyCentralAndeanMantle2014}.
The absolute depth of the crustal seismicity towards the east is likely overestimated, due to their location outside the network and near-vertical ray departure angles.

Second, we observe seismicity clusters in the upper plate starting right above the interface at shallow interface depth ($<$45~km) covering predominantly a vertical range of up to $\sim$ 10~km (Figure~\ref{fig:upper_plate}, cyan to green colors).
These clusters occur at distances of up to 1\degree ~from the trench and are predominantly visible in three areas (Figure~\ref{fig:cross_section_ortho}): around Taltal (25\degree S to 26\degree S), around Copiapó (27\degree S to 28\degree S), and around La Serena (29\degree S to 30\degree S).
The seismicity delineates the offshore continuation of Atacama fault system mapped onshore \cite{seymourMagnitudeTimingRate2021,maldonadoComprehensiveDatabaseActive2021}.
This is particularly clear in the Los Vilos - La Serana - Los Choros segment (Figure~\ref{fig:upper_plate}, 28.5\degree S to 30.5\degree S).
Furthermore, two of the segments (Taltal and Copiapó) are located where an oceanic seamount chain enters in subduction.
All three areas are located between highly locked asperities, and correspond to a local low in interseismic coupling.
Similar links between coupling degree, shallow seismic activity, and topographic features have previously been proposed \cite{bejar-pizarroAndeanStructuralControl2013,saillardSeismicCycleLongterm2017,oryanMegathrustLockingEncoded2024}.
The cross-sections in Figure~\ref{fig:cross_section_crustal} suggest that the cluster in the Taltal region is connected to the surface through splay structures.

Comparing the locations of these events with the Moho model from \citeA{rivadeneyra-veraUpdatedCrustalThickness2019}, it is not immediately clear if these events locate in the continental mantle or crust.
Nonetheless, several factor point at crustal activity.
First, we observe a further cluster of seismicity downdip, with a clear separation, providing evidence for the location of the mantle wedge corner.
A similar double structure of crustal and mantle wedge seismicity has been reported further south in the aftershocks of the 2010 $M_w=8.8$ Maule earthquake \cite{wangRoleSerpentinizedMantle2020}.
Second, in our 3D velocity model, we observed lower P wave velocities, typical for the crust, in the area of these clusters while observing higher velocities just downdip (Figures~\ref{fig:velocity_dvp}).
Third, as previously mentioned, the seismicity aligns well with mapped surface faults (Figure~\ref{fig:upper_plate}).
This seismicity closely mirrors the crustal wedge seismicity detected further north along the subduction zone, in particular, in the area between 21\degree S and 21.5\degree S \cite{sipplSeismicityStructureNorthern2018}.
Based on the thermal model from \citeA{wadaCommonDepthSlabmantle2009}, \citeA{sipplSeismicityStructureNorthern2018} suggest that this activity only occurs at temperatures below 350-400\degree C, the temperature range where quartz becomes ductile.

Third, we observe seismicity in the mantle wedge (Figure~\ref{fig:upper_plate} yellow to red), consisting of deeper clusters, occurring at 1.2\degree ~to 1.6\degree ~from the trench (e.g. Figure~\ref{fig:cross_section_ortho}, 28.0\degree S - 28.5\degree S).
This group has an almost vertical cutoff towards the east above an interface depth of roughly 65~km.
It starts at the interface and extends up to 30~km vertically upwards.
Comparing the location of the seismicity to the continental Moho model from \citeA{rivadeneyra-veraUpdatedCrustalThickness2019}, it can be clearly identified that this seismicity locates in the mantle wedge.
Thermal models around Antofagasta to the north and Illapel in the south confirm that the observed seismicity coincides with the decoupled mantle wedge \cite{wangSoftBarrierMegathrust2025}.
The decoupling from the mantle flow causes the lower temperature of this part of the mantle, known as the cold nose, supporting the mantle wedge seismicity \cite{wadaCommonDepthSlabmantle2009,peacockStabilityTalcSubduction2021}.
The mantle wedge seismicity is clearly separated from the crustal seismicity along dip by a seismically quiet section of the upper plate, covering at least 0.2\degree ~(e.g. Figure~\ref{fig:upper_plate}, 29.5\degree S - 30.0\degree S).

Mantle wedge seismicity only occurs south of 27\degree S, coinciding with the flat slab, even though updip the flat segment, and the volcanic gap.
Variations in mantle wedge seismicity along strike are more smooth than for the more localized crustal seismicity.
The absence of mantle wedge seismicity north of 27\degree S is consistent with its absence between 18\degree S and 24\degree S \cite{sipplNorthernChileForearc2023}.
While for the crustal wedge seismicity, we observe a clear relationship between activity rates and interplate locking, the reasons for along-strike variation in mantle-wedge seismicity are less clear.
A potential explanation is a different degree of serpentinization of the oceanic plate \cite{halpaapEarthquakesTrackSubduction2019}.
Mantle wedge seismicity is typically linked to serpentinization due to fluid intrusion into the cold nose of the mantle wedge \cite{halpaapEarthquakesTrackSubduction2019,angiboustJoltsJadeFactory2021}.
A key factor for the serpentinization of the oceanic plate is outer rise seismicity \cite{hatakeyamaMantleHydrationOuterrise2017}, in particular, in a sediment-starved subduction such as northern Chile.
Indeed, we find an increased rate of outer rise seismicity in the southern part of the study area, corresponding with the mantle wedge seismicity.
This fluid is then released in dehydration reactions in the subducting plate, before ascending into the mantle wedge \cite{halpaapEarthquakesTrackSubduction2019}.
The highly active intraslab seismicity just downdip the imaged mantle wedge seismicity provides ample evidence for such dehydration reactions in the slab.

\subsection{Outer rise seismicity}

The study area shows active outer rise seismicity.
Between 24\degree S and 27\degree S, outer rise seismicity is concentrated in one band to the west of the trench.
Similarly, south of 30\degree S, outer rise seismicity is compressed into a single band, in this case near the trench.
Between these two segments, in the area from 27\degree S to 30\degree S, a wider band of outer rise seismicity is observed that splits into multiple lineaments (Figure~\ref{fig:outer_rise}).
These lineaments are sub-parallel to the trench, with a strike around 30\degree .
They are consistent with fault traces in the bathymetry, particularly visible in the area between 28\degree S and 29\degree S around 72.5\degree W.
This seismic activity is related to the curvature of the trench, leading to the bending of the oceanic plate before it subducts \cite{geersenDoesPermanentExtensional2018}.
Outer rise activity levels vary along strike, with event density north of 27\degree S, where plate coupling is high, only about half the rate to the South, where plate coupling is low.
This suggests a potential link between locking degree and outer rise activity \cite{christensenSeismicCouplingOuter1988}.
As noted before, the depth of the outer rise seismicity is likely overestimated due to the poor azimuthal coverage.

\section{Time-space segmentation of the subduction interface}

\subsection{Interface seismicity and its large-scale segmentation}

\begin{figure*}[ht!]
\centering
\includegraphics[width=\textwidth]{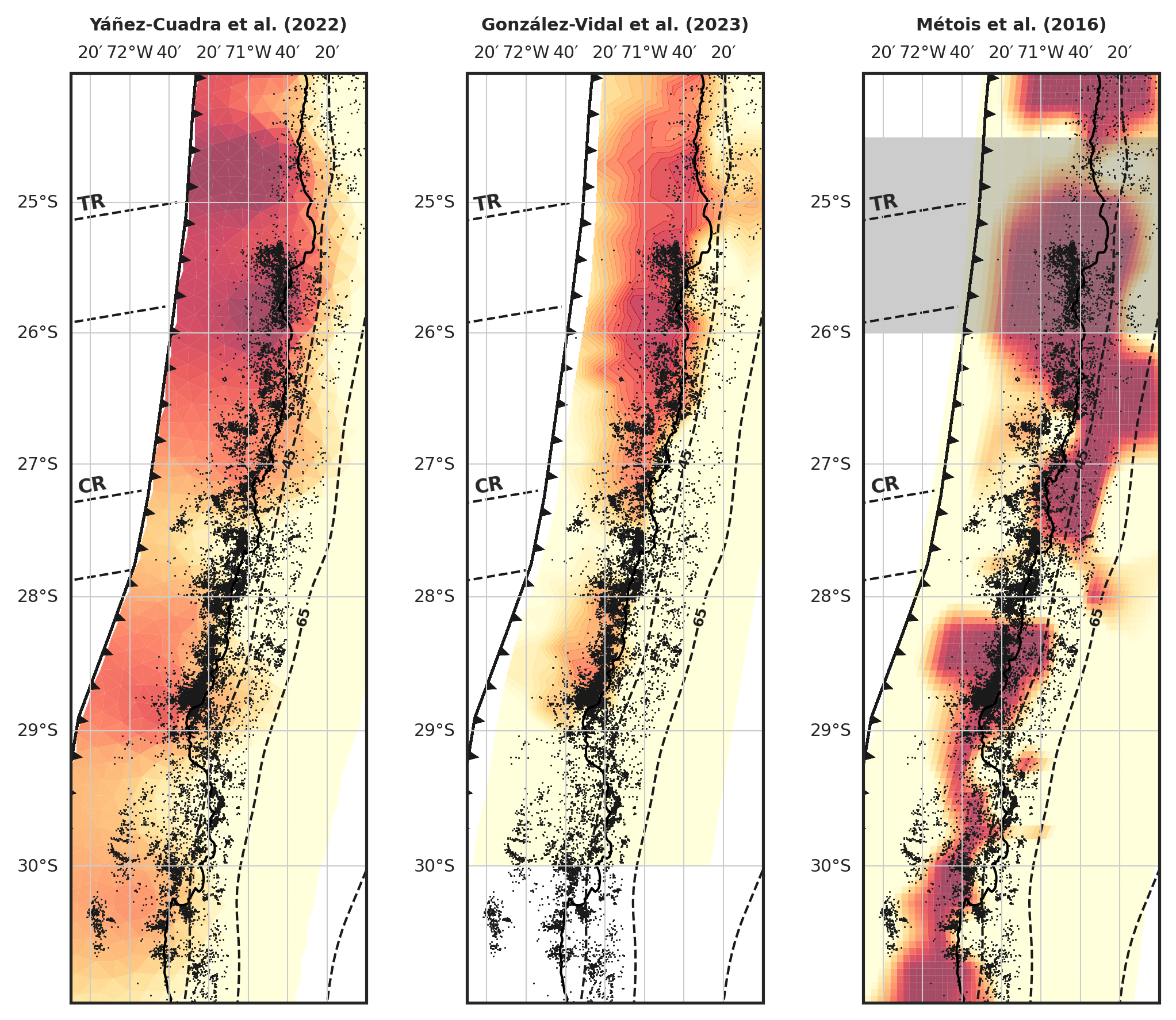}
\caption{Map view of the interface seismicity in comparison with different interplate locking models. Black dashed lines indicate depth contours of the slab. The interplate locking models are from \citeA{yanez-cuadraInterplateCouplingSeismic2022}, \citeA{gonzalez-vidalRelationOceanicPlate2023}, and \citeA{metoisInterseismicCouplingMegathrust2016}. TR and CR denote the incoming Taltal and Copiapó ridges. Note that no locking data is available south of 30\degree S for the model of \citeA{gonzalez-vidalRelationOceanicPlate2023} and that \citeA{metoisInterseismicCouplingMegathrust2016} indicate an area of low resolution between 24.5\degree S and 26\degree S (grey overlay).}
\label{fig:interface_seismicity}
\end{figure*}

The largest seismic moment release on long time scales happens on the subduction interface.
While our catalog does not contain a large megathrust earthquake, we detect a total of 36,165 interface earthquakes.
Figure~\ref{fig:interface_seismicity} shows the interface seismicity overlain on  geodetically inferred interplate locking models.
The majority of interface seismicity occurs in a band of around 50~km width updip the 45~km isodepth line of the slab surface.
Between 45~km and 65~km interface depth, events occur more isolated or in less active clusters.
As an exception to this, south of 29\degree S, interface seismicity exhibits a wider spread along dip, with dense seismicity clusters extending down to 55~km interface depth.
At a depth of around 60~km, interface seismicity arrests, marking the transition from brittle failure to ductile deformation on the interface \cite{wadaCommonDepthSlabmantle2009}.
In regions with seismicity in the mantle wedge, this transition coincides with the termination of the deep upper plate activity.

The interface seismicity distribution is highly localized and appears in patches.
The patches show varying shapes, including circular, narrow linear or arc-shaped patterns, with extents from few kilometers up to almost 50~km.
Patches have a narrow vertical extent (generally below 1~km) and are oriented parallel to the slab surface.
The overall activity shows strong variation along strike.
The highest activity occurs between 30\degree S and 27\degree S, but activity rates vary strongly over short spatial scales.
For example, between 28.7\degree S and 28.8\degree S we observe more than 3,500 events while only around 300 events were recorded just 50~km further north.
This stands in contrast to the intraslab seismicity which shows a much smoother along-strike variation in activity levels.

Figure~\ref{fig:interface_seismicity} compares the interface seismicity to interplate locking model inferred from geodetic records.
We observe that regions with low geodetic locking are characterised by seismic activity extending substantially further towards the trench than highly locked regions.
Examples are the area of the Copiapó ridge around 27.5\degree S and the area around 29.5\degree S.
Around the Copiapó ridge, the seismic activity occurs in swarms indicating shallow SSE activity \cite{munchmeyer2024chile_sse,ojedaSeismicAseismicSlip2023,marsanEarthquakeSwarmsChilean2023} that at the same time contributes to the low coupling.
No similar characterization is available for the segment around 29.5\degree S.
A connection between low locking and enhanced interface seismicity has previously been observed in central and northern Chile \cite{sipplMicroseismicityAppearsOutline2021,gonzalez-vidalRelationOceanicPlate2023}.
Downdip, interface seismicity activity reduces with the reduced interplate coupling.
The more isolated seismicity patches in this area correspond well with the Domain C proposed by \citeA{layDepthvaryingRuptureProperties2012}, an area where the seismic patches occur in isolation on the predominantly non-coupled interface.

It is likely that interplate locking and the seismicity distribution correspond on smaller scales too, forming small-scale Mogi doughnuts on the downdip edge of highly coupled patches \cite{kanamoriNatureSeismicityPatterns1981}.
However, the difference in resolution between seismic (below 100~m) and geodetic (tens of kilometers) observables makes a clear conclusion difficult.
In addition, geodetic inversions have lower temporal resolution than seismic observations, making it difficult to identify short-term transient changes.

\subsection{3D structure of the subduction interface}
\label{sec:interface}

\begin{figure*}[ht!]
\centering
\includegraphics[width=\textwidth]{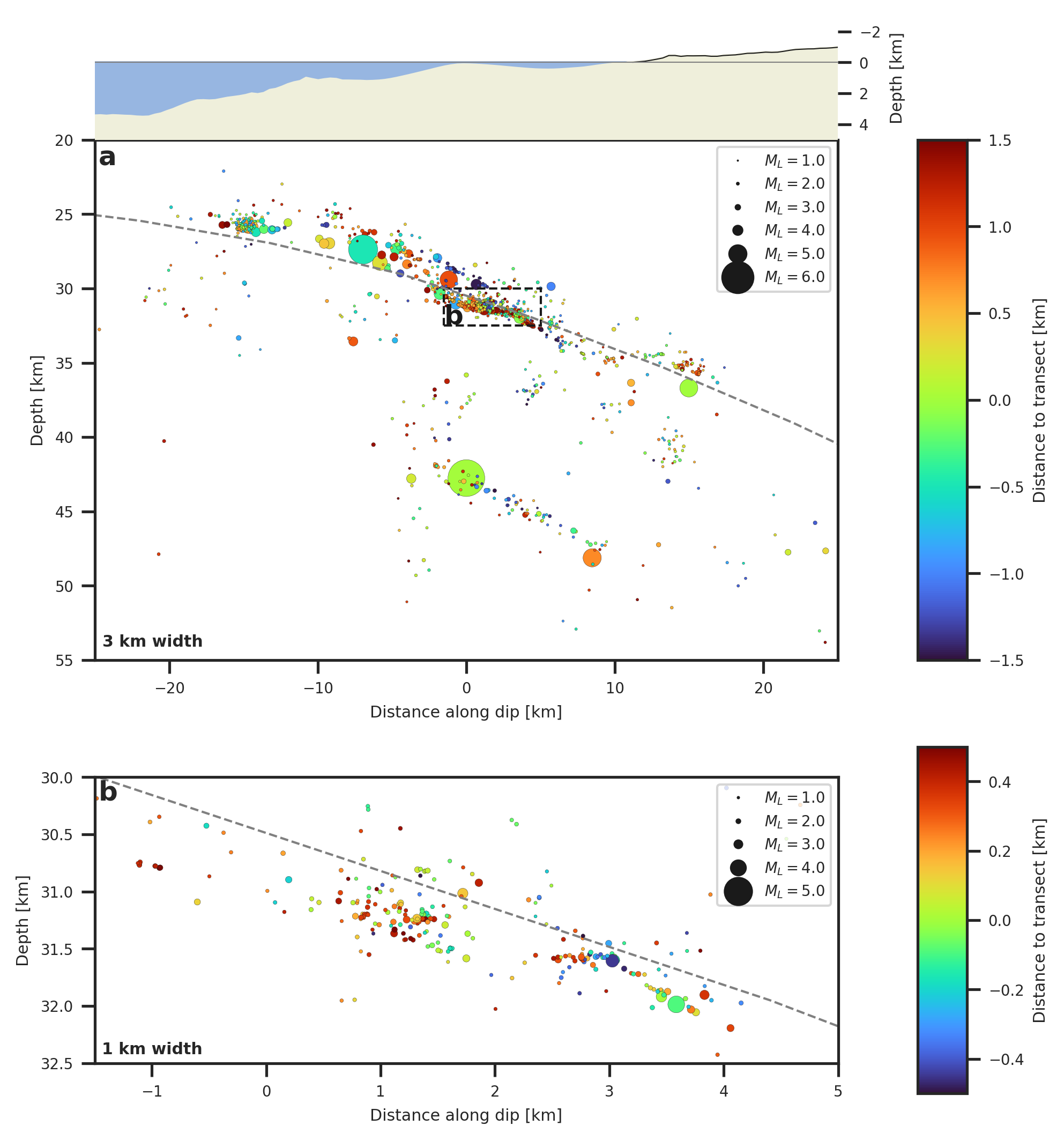}
\caption{Cross-section through an active cluster along 28.77\degree S. The 0~km mark is set at 71.48\degree W. \textbf{a} Cross-section of interface and intraslab events with a width of 3~km, i.e., 1.5~km to each side of the cross-section center. The overlying topography and bathymetry is displayed on top. \textbf{b} Zoom-in on the most active interface cluster (black rectangle in a) with a width of 1~km. Each dot represents one earthquake, scaled by magnitude and colored by the orthogonal distance to the transect center. To highlight fine-scale differences, we use an a zoomed-in color scale for panel b. The grey dashed line indicates the slab model we inferred from the data. The small disagreement between the interface visualised and the interface seismicity originates from the smoothing of the slab model.}
\label{fig:cross_section_example_combined}
\end{figure*}

Recent research shows that dense seismic networks and modern processing techniques enable us to resolve the 3D structure of the subduction interface \cite{chalumeauSeismologicalEvidenceMultifault2024}.
Figure~\ref{fig:cross_section_example_combined} show an example cross section from our catalog.
In Figure~\ref{fig:cross_section_example_combined}a, we analyse a 3~km wide swath of seismicity, showing both interface and intraslab events.
The subduction interface can clearly be identified by a band of seismicity.
The width of this band varies from about 500~m to 2.5~km, providing an estimate for the width of the interface.
Notably, we can identify substructures within the interface.
These structures are generally parallel to the interface, even though with slight variations in dip.
Different parallel structures are offset in trench-perpendicular direction, i.e., they do not occur directly on top of each other.
Our slab model (grey line) shows a good fit with the interface seismicity, yet fails to cover the full detail due to the imposed smoothness and the simplification of the interface to a 2D surface.

Looking at the lower boundary of the interface seismicity, we identify some abrupt changes, such as the one highlighted in the black box, where seismicity switches from a more oblique to a steeper orientation with a clear corner.
Figure~\ref{fig:cross_section_example_combined}b shows a zoom in of this feature with a section width of 1~km.
Two lines of seismicity, each with vertical extent around 50~m and along dip extent below 1~km delineate the change in seismicity.
While the updip line is oriented horizontally, the downdip line dips at an angle of $\sim30$\degree.
Further updip, seismicity does not show similarly clear lines but is more diffuse within a band of $\sim500$~m width.
Looking back at Figure~\ref{fig:cross_section_example_combined}, this band widens to about 2.5~km kilometer and persists around 15~km updip, before reducing down to around 500~m again.

The structures imaged might indicate underplating and subduction erosion processes \cite{cliftSlowRatesSubduction2007,kimuraSeismicEvidenceActive2010,schollSeismicImagingEvidence2021}.
Such processes might correspond to forearc uplift, consistent with the observed topography \cite{menantTransientStrippingSubducting2020,cubasEarthquakeRupturesTopography2022}.
This is consistent with their location underneath the coastline and in the area of partial interplate coupling.
The cause of this underplating can both be the general change in slab geometry, i.e., the visible steepening of the slab, or the effect of subducted topography inducing additional fractures around the plate interface \cite{mochizukiWeakInterplateCoupling2008}.

\subsection{Fine-scale interface segmentation}
\label{sec:segmentation}

\begin{figure*}[ht!]
\centering
\includegraphics[width=0.8\textwidth]{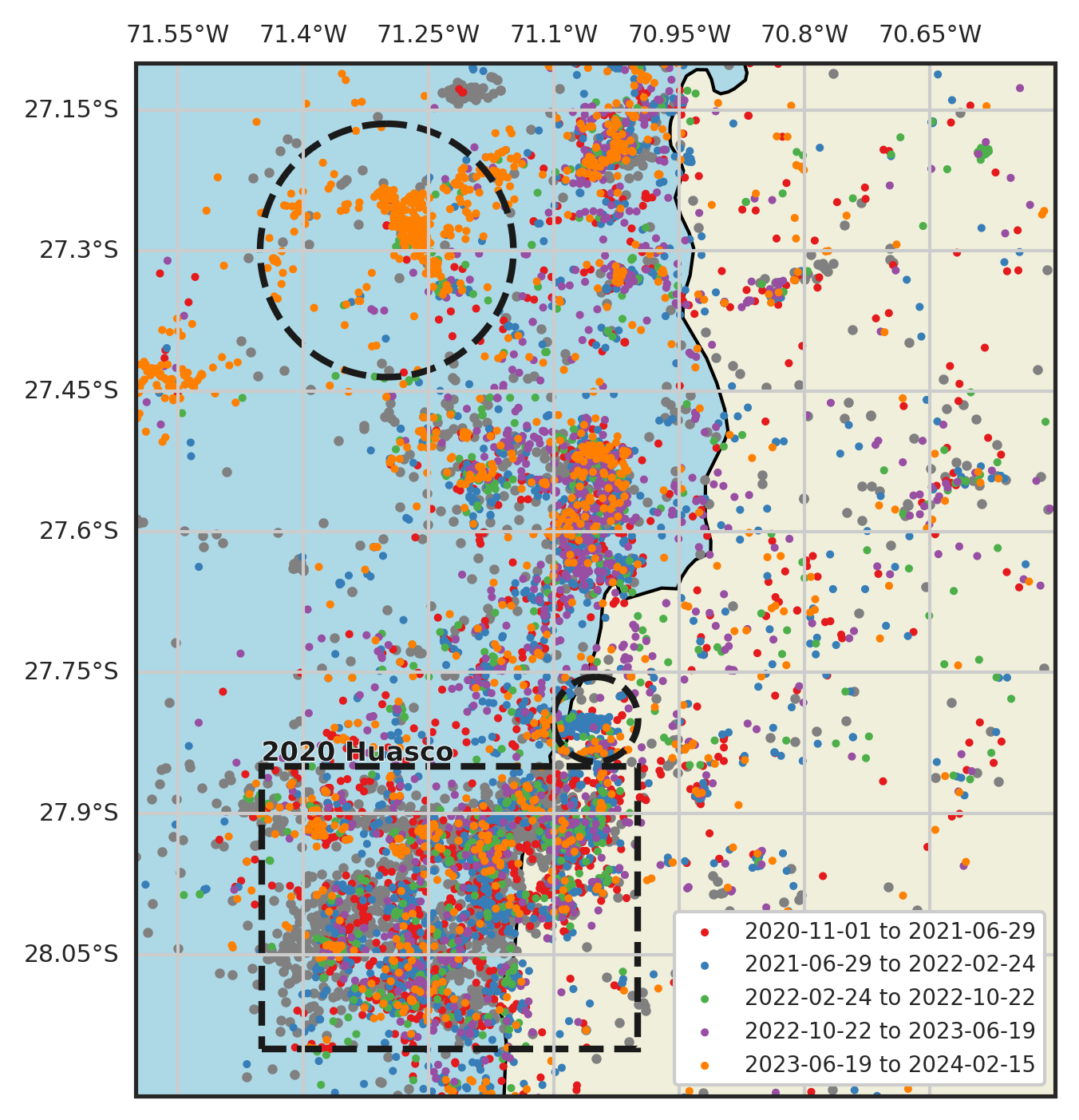}
\caption{Interface activity in the area of the subducting Copiapó ridge and the 2020 Huasco earthquake. The activity is split into 5 disjoint segments by time. Each segment is shown in a different color, with the order of events randomized. Below, we plot events from the reference catalog covering 2014 to 11/2020 in grey. The reference catalog uses the same 3D velocity model but has not been relocated jointly with the main catalog. It relies on permanent stations only and is therefore less complete and has higher location uncertainties. Black dashed circles indicate example areas with activity almost exclusively during a single time block. The black rectangles highlights the 2020 Huasco sequence.}
\label{fig:segmentation_stability}
\end{figure*}

\begin{figure*}[ht!]
\centering
\includegraphics[width=\textwidth]{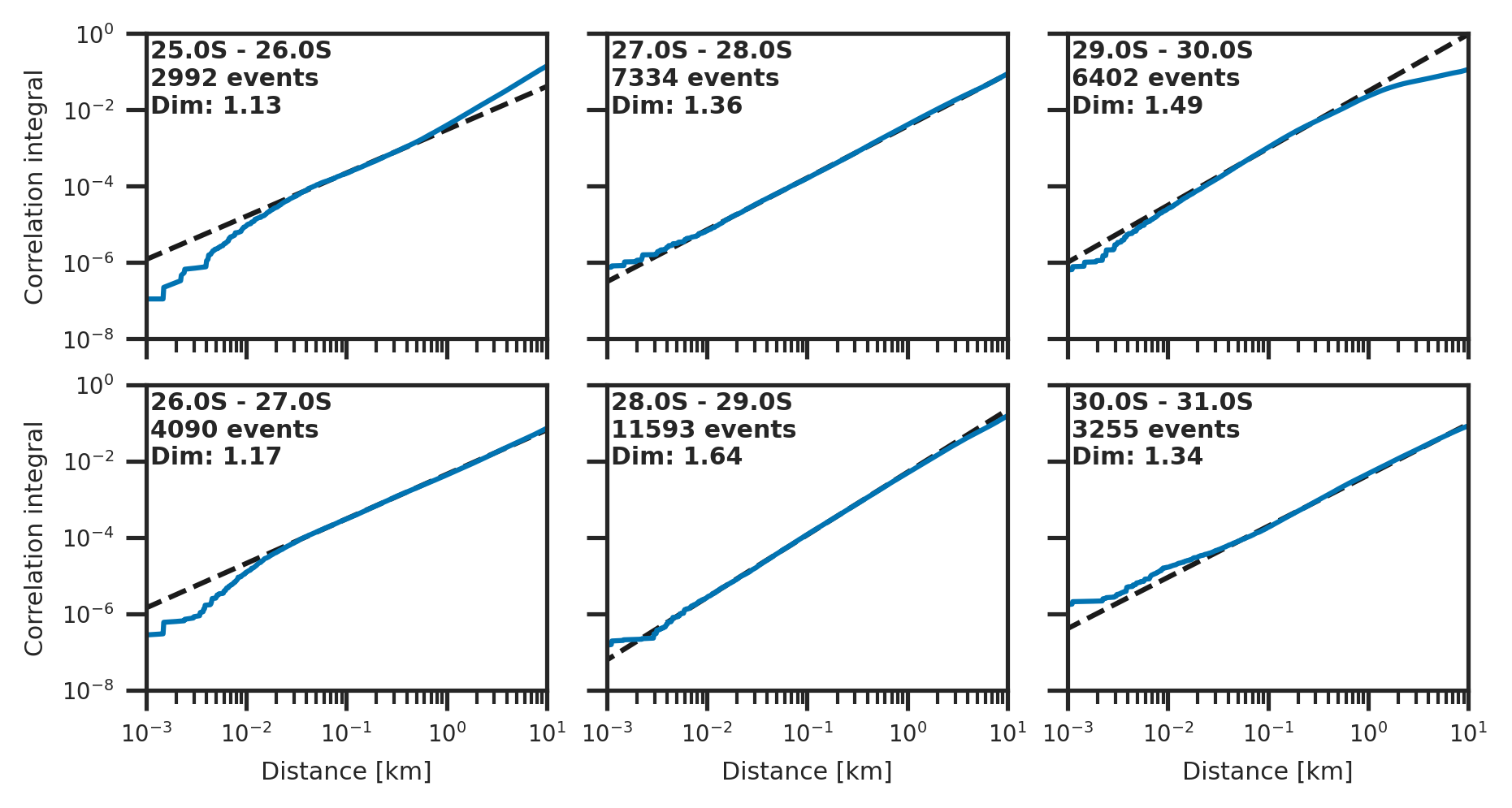}
\caption{Correlation integral for interface events grouped by latitude. In the upper left corner, we provide the latitude range, the number of events, and the estimated fractal dimension. We estimate the fractal dimension using a log-log fit between 30~m and 500~m, as we observed consistent power law behaviour in this range. We calculate the correlation integral from the epicentral positions, i.e., disregarding the event depth. See Figure~\ref{fig:interface_fractal_3d} for an estimation using hypocentral locations.}
\label{fig:interface_fractal}
\end{figure*}

After studying the large-scale segmentation and fault zone width of the interface, we now turn towards the fine-scale segmentation.
Several studies have proposed a segmentation of the interface into seismic and aseismic area \cite{senoFractalAsperitiesInvasion2003,perfettiniModelAftershockMigration2018,behrTransientSlowSlip2021}.
This is consistent with our finding of clustered seismicity among the interface events.
To study this phenomenon in further detail, we consider two aspects: the temporal stability of the segmentation and the spatial scale of the segmentation.
Figure~\ref{fig:segmentation_stability} (colored dots) shows the interface activity in the region of the subducting Copiapó ridge and the $M_w=7.1$ 2020 Huasco sequence \cite{molina2024sse}, grouped into different time periods.
We chose this section as it exhibits high seismic activity and had good station coverage throughout the study duration.
Comparing the different periods, it is clear that most clusters are active across almost all periods.
This holds both for highly active clusters, such as in the aftershock area of the 2020 Huasco earthquake, and for less active clusters.
At the same time, a few clusters are almost exclusively active during one time period.
We highlight two such examples with black, dashed circles in Figure~\ref{fig:segmentation_stability}.
We suggest that these events are caused by an outside driver.
For example, the region hosting the orange cluster in Figure~\ref{fig:segmentation_stability} is known for hosting seismic swarms related to shallow slow slip \cite{ojedaSeismicAseismicSlip2023,molina2024sse,munchmeyer2024chile_sse}.

To extend the temporal duration of our analysis, we created a second catalog, based exclusively on the permanent seismic stations available in the region.
To ensure comparable locations, we use the same 3D velocity model and station residuals to locate the events in this catalog.
However, relative relocation is performed independently, i.e., comparability of locations is limited by the precision of the absolute locations.
While the catalog is not as complete as our permanent catalog, it allows us to compare the activity to a longer reference frame.
Figure~\ref{fig:segmentation_stability} (grey dots) shows that the interface segmentation is stable over the full 10 year duration from 2014 to 2024.
The seismically active regions in the 10 year catalog almost completely encompass the seismic activity of the 3.5 year catalog, even though the latter is substantially more complete.
At the same time, only few events in the longer duration catalog occur in regions that do not show seismic activity between 11/2020 and 02/2014.

To study the spatial scale of segmentation, several studies analysed fractal dimension of the seismicity distribution \cite{akiProbabilisticSynthesisPrecursory1981,turcotteFractalsChaosGeology1997,harteDimensionEstimatesEarthquake1998,kaganEarthquakeSpatialDistribution2007}.
The fractal dimension can be estimated from the correlation integral.
For a given distance $d$, the correlation integral is given by the fraction of event pairs with an interevent distance below $d$.
For a fractal distribution, i.e., a distribution that is self-similar across different scales, the correlation integral has a power-law fit with the distance.
The exponent of this power law fit is called the fractal dimension and describes the distribution of the events.
A low fractional dimension suggests high clustering of the events in space, while a high value corresponds to a more uniform distribution.
Using epicentral locations, for subduction zones \citeA{harteDimensionEstimatesEarthquake1998} inferred fractal dimensions between 1.2 (shallow events in Kanto) and 1.8 (deep events in New Zealand).

To study the interface structure, we estimate correlation integrals and fractal dimensions grouped by latitude (Figure~\ref{fig:interface_fractal}).
Along the subduction zone, we observe stable fractal behaviour for length scales between 30~m and 500~m that we use to estimate fractal dimensions.
However, fractal behaviour is stable over larger distance ranges for some parts of the interface: between 26\degree S and 27\degree S up to more than 10~km, between 27\degree S and 29\degree S from below 10~m up to over 10~km.
This scaling, in some cases over more than three decades in length, shows that the subduction interface is not only separated into seismic and aseismic patches, but that the seismic patches themselves are actually subdivided in a fractal way.
Depending on the size of an earthquake, it will rupture patches on different scales, with aseismic areas in between dense seismic patches being broken through in a dynamic manner.
The fractal dimensions show a strong but smooth variation along strike, with values from 1.13 (25\degree S to 26\degree S) to 1.64 (28\degree S to 29\degree S).
Notably, the highest values fractal dimensions, i.e., the most diffuse distribution of events is observed in the area of the Huasco earthquake 2020, that occurred about 3 months before the beginning of our catalog.
It is unclear if the higher fractal dimension is a result of the aftershock activity, or if it corresponds to an interface structure that enabled the nucleation and propagation of this larger event.

In addition to describing the clustering of the events, the fractal dimension provides an estimate of the relative location errors for interface events.
We can estimate an upper bound on the relative error from the lower length scale at which the correlation integral deviates from the power law fit.
As expected, we find higher uncertainties for events at the boundaries of the network ($\sim$30~m) than within the network (down to 10~m).
However, we note that such levels of precision are only achieved among closely colocated events, as these events feature highly similar waveforms, allowing for accurate differential travel-time estimation.
Nonetheless, it highlights the excellent relative location quality that can be achieved even for interface events with a dense, deep learning catalog.

\subsection{Recurrence times and temporal clustering of events}

\begin{figure*}[ht!]
\centering
\includegraphics[width=0.7\textwidth]{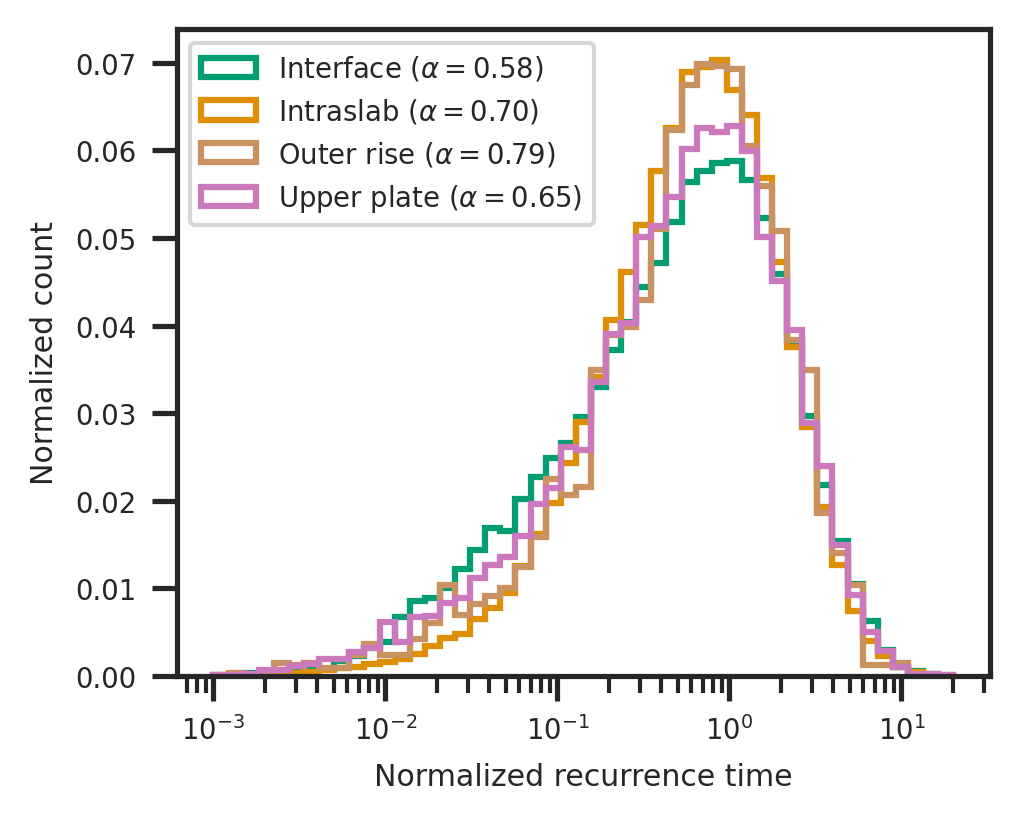}
\caption{Distribution of recurrence times by event category. For each category, we determine all inter-event times of consecutive events. To account for the different number of events between the categories, we normalize by dividing by the average recurrence time. We scale the histogram by the total event count to obtain comparable curves. The numbers in brackets provide the scale parameter $\alpha$ for a Gamma distribution.}
\label{fig:recurrence}
\end{figure*}

We compare the recurrence intervals between the different categories of events (interface, intraslab, outer rise, upper plate).
We exclude mine blasts and airgun shots from this analysis due to their clear anthropogenic signature.
Figure~\ref{fig:recurrence} compares the distribution of normalized recurrence times.
All categories have shapes typical for the Gamma distributions of earthquake interevent times \cite{corralLocalDistributionsRate2003}.
Nonetheless, the distributions vary with regards to their scale parameter $\alpha$, i.e., the ratio of shorter inter-event times among all times or, in other words, the degree to which the event distribution is burst-like.
We estimate the shape parameters $\alpha$ from the empirical mean $\mu$ and standard deviation $\sigma$.
\begin{equation}
 \alpha = \left( \frac{\mu}{\sigma} \right) ^2
\end{equation}
We observe the lowest shape parameters for interface events ($\alpha=0.58$) followed by upper plate events ($\alpha=0.65$).
This highlights that these two event classes are characterised by more burst-type occurrences.
In contrast, for intraslab events ($\alpha=0.70$) and outer rise events ($\alpha=0.79$), we observe substantially higher shape parameters, indicating that these classes are less burst-like, i.e., more uniformly distributed over time.
The findings are mirrored visually in the histograms of Figure~\ref{fig:recurrence}, where interface and upper plate events show a substantially higher number of short recurrence times than intraslab or outer rise events.
Similar behaviour has previously been reported for Northern Chile \cite{sipplSeismicityStructureNorthern2018}.
\citeA{cabreraNorthernChileIntermediatedepth2021} suggest that this difference can be explained by different aftershock productivity, which is thermally controlled.
They suggest that the 400 to 450\degree C isotherm, running close below the interface, separates zones of low and high hydration, thereby causing the contrast in aftershock productivity.
More broadly, research across different subduction zones suggests lower aftershock productivity with higher depth and for intraslab events \cite{gombergProductivityCascadiaAftershock2021,chuAftershockProductivityIntermediatedepth2022,wimpennyReExaminingTemporalVariations2023}.

The findings on temporal clustering mirror the previously reported findings on spatial clustering.
Classes of events that are characterized by a high spatial clustering, such as interface events, are also characterised by a high temporal clustering.
This suggests that in addition for thermal control, the temporal clustering might be related to the structure of the seismogenic zones.
Classes with high spatial clustering, such as interface events, exhibit low distances between potentially seismogenic area.
This means that the stresses transferred between events and potentially seismogenic areas are higher, leading to an enhanced number of triggered seismicity, manifesting in a higher number of short inter-event times.

\section{Conclusion}

In this study, we compiled a dense, high-resolution seismicity catalog for the Atacama segment in Northern Chile.
We include waveforms from almost 200 seismic stations from temporary and permanent deployments.
Using a deep learning workflow, we detect more than 165,000 events within 3.5 years.
For regions within the study area, we report relative location uncertainties well below 100~m and a magnitude of completeness of 1.5 or lower.

The seismicity in the Atacama segment can be grouped into several structures.
West of the trench, we observe outer rise events that align with seafloor bathymetry.
On the plate interface, we observe seismicity predominantly down to 45~km interface depth.
The largest number of events occurs within the subducting slab.
In most parts of the region, this intraslab seismicity consists of two slab-parallel layers.
In addition, we observe three seismicity clusters in the deep Pipanaco nest and one cluster in the deep Jujuy nest.
The upper plate shows seismicity on shallow crustal fault systems, in a crustal wedge, and in the mantle wedge.
Crustal seismicity correlates with the slab geometry, with a narrow extend in the area of the dipping slab and a wide band of seismicity in the area of the flat slab.
Compared to the adjacent segment in the north, the seismicity in the Atacama segment shows more along-strike variation: this includes the changing separation of the intraslab seismicity planes and the mantle wedge seismicity.
This higher variability is likely caused by the more complex geometry with the changing strike of the trench and the flat slab segment.

We used our catalog to infer structural properties.
First, we analysed the slab geometry and showed that the extent of the flat segment is narrower than previously imaged.
Furthermore, the flat slab is dipping towards the south.
Second, the distribution of seismicity reveals the fine structures of the subduction interface.  
For example, in cross-sections, the interface width varied between several hundred meters and 2.5 kilometers.
Seismicity follows a fractal distribution over three orders of magnitude, suggesting a fractal distribution of seismic patches within an aseismic matrix.
This segmentation is stable at least on the scale of a decade, as shown with an longer duration catalog based on permanent seismic network data.

In this study, we provide an overview of the dense seismicity in the Atacama segment in Northern Chile.
Our catalog improves the understanding of this mature seismic gaps, and the processes underlying the complex seismic and aseismic processes hosted in the region.
In addition, we make our catalog and related data products available.
We hope our catalog can serve as a basis to understand this major seismic gap with its complex interaction of slow and fast deformation.

\section*{Open Research}
We publish our catalog with the associated phase picks, the longer duration catalog based on permanent stations with the associated phase picks, the 1D and 3D velocity models, and the slab model.
A publication with Zenodo is in preparation.

\acknowledgments
This work has been partially funded by the European Union under the grant agreement n°101104996 (“DECODE”) and the ERC CoG 865963 DEEP-trigger. Views and opinions expressed are however those of the authors only and do not necessarily reflect those of the European Union or REA. Neither the European Union nor the granting authority can be held responsible for them.
We thank the RESIF, GEOFON and Earthscope data centers for making data available.
We thank the GFZ Potsdam GIPP and the French national pool of portable seismic instruments SISMOB-RESIF (INSU-CNRS) for providing seismological instruments and related metadata used in this study.
We thank INPRES for making the data from the RI network available to us.
We thank INPRES and CIGEOBIO (Universidad Nacional de San Juan) for the data from the stations DOCA and PEDE.
Stations from the 9C network were funded by CNRS-Tellus "COP2020: SlowSlip Trigger" and ANR-S5 (ANR-19-CE31-0003) projects.
D.M.'s salary was covered by a grant from CNES.
J.-C.B. and M.M. acknowledge funding from ANID PIA-ACT192169 Anillo PRECURSOR project.
J.-C.B. acknowledge support from FONDECYT 1240501 ANID project.
DL thanks the Bundesministerium für Bildung und Forschung (BMBF) for support (grant 03G0297A).
M.M. acknowledges support from ICN12019N Instituto Milenio de
Oceanografía and FONDECYT project 1221507.
We thank everyone involved in the deployment of the seismic and geodetic networks and the data management.
We thank the hosts and landowners of all station sites.
The computations presented in this paper were performed using the GRICAD infrastructure (\url{https://gricad.univ-grenoble-alpes.fr}), which is supported by Grenoble research communities.

\bibliography{mybibfile,zotero}

\clearpage
\appendix

\section*{Supplementary texts}

\subsection*{Text S1 - Comparison to reference catalogs}

We compare our catalog to the routine catalog generated by the Chilean Seismic Network (CSN) based on permanent seismic stations.
The CSN catalog contains $\sim$6100 events in the study region during our catalog period, less than 4~\% of the events in our catalog.
We match events between the two catalogs based on origin times.
Figure~\ref{fig:csn_comparison} shows a comparison of the event locations and magnitudes.
Overall, location differences within the network are in the range of few kilometers.
A similar scatter can be observed for the depth within the network.
Only minor systematic differences are visible here.
Outside the network, differences between the catalogs are larger.
Offshore, the CSN catalog consistently reports shallower depth than our catalog, consistent with the observation that we overestimate the depth of outer rise events.
Epicentral locations show higher scatter than onshore but no systematic bias.
As our locations exhibits more clustering than the CSN catalog, this scatter likely results from uncertainties in the CSN catalog.
Towards the east, systematic differences in depth are low, however our catalog consistently locates events further east than the CSN.
For the deep Jujuy cluster around 24\degree S, we locate events on average 40~km deeper than the CSN.

The systematic location differences result from different velocity models employed for locating the events.
As we inverted our velocity model based on the dense station coverage and seismicity, we suggest that our absolute location are more accurate than the ones reported by CSN.
In particular, the inclusion of four stations in Argentina provides us with better coverage on the deep events towards the East of our study area.
Similarly, our dense station coverage enables us to reduce the uncertainties on the absolute and relative locations.
This suggests that most scatter between the catalogs results from the uncertainties in the CSN catalog.

For magnitudes, we observe a good agreement between the two catalogs.
Due to the normalisation in our magnitude estimation procedure, no systematic offset exists between the scales.
The standard deviation between the catalogs is 0.21 magnitude units.
It results from variations in the amplitude measurements and differences in the attenuation function employed.
The magnitudes scale 1:1 with each other, with the exception of events above magnitude 6, which are systematically underestimated in our catalog.
This is most likely caused by including clipped records in the calculation of our magnitude values, i.e., underestimating amplitudes at stations with low epicentral distance.

As an additional validation, we compare our catalog to the GCMT catalog \cite{ekstromGlobalCMTProject2012}.
Figure~\ref{fig:cmt_comparison} shows the hypocenter and magnitude comparison.
The results are consistent between both catalogs within few kilometers in both epicentral and depth estimates.
On average, GCMT provides locations slightly further offshore than our catalog.
While we observe no systematic difference in depth onshore, towards the trench GCMT depth are generally more shallow than our estimate.
As these shallower estimates are compatible with the bathymetry at the trench, this suggests that our offshore locations overestimate depth due to the limited azimuthal coverage.
Compared to the $M_w$ values from $GCMT$, our local magnitudes are on average 0.10 magnitude units larger.
The scatter is 0.21 magnitude units.

\subsection*{Text S2 - Estimating magnitude of completeness}
\label{sec:supp:completeness}

To estimate $M_c$ and its spatial variation for the different deployment periods, we use a modified version of the method by \citeA{schorlemmerProbabilityDetectingEarthquake2008}.
The method relies on estimating detection probabilities with respect to magnitude and hypocentral distance.
These detection probabilities are estimated individually at each stations to account for station conditions, such as instrument, local geology, and typical noise conditions.
From the individual station results, for each location and magnitude, detection probabilities can be inferred with respect to each possible station distribution.
As the method is event-centered, it does not required averaging over time periods or areas.
Therefore, it is particularly useful for catalogs with highly varying station configurations as in our study.

Compared to the original method of \citeA{schorlemmerProbabilityDetectingEarthquake2008}, we used several modifications.
As triggering criterion, we use whether a station has both associated P and S picks as this is the most strict criterion in our quality control.
Instead of analytically estimating detection probabilities for events, we use Monte Carlo sampling.
An event is successfully detected if at least 5 stations have both P and S detections.
For consistency, we use the same Monte Carlo samples for each target magnitude.
We define $M_c$ as the smallest magnitude with detection probability above 0.999.
As detection probability curves are steep, a higher threshold would not substantially alter $M_c$.
However, it would be numerically more expensive due to the Monte Carlo sampling.
For estimating attenuation, we use the curves inferred in the magnitude estimation step.

The method assumes independence between detection probabilities at different stations.
While correct to first order, especially closely spaced stations will have correlated detections due to, e.g., noise conditions or source effects.
This can lead to underestimations of $M_c$ in regions with dense spacing.
In addition, the method has a bias for very small events due to sampling bias.
To mitigate this bias, we do not infer detection probabilities if less than 8 events within 0.1 magnitude units of a magnitude-distance pair have been observed, i.e., we set detection likelihoods for these magnitude-distance pairs to zero.
The rational is that at such low magnitudes, a much higher count of detections is expected, i.e., the apparent high detection probability is only an artifact of the sampling bias.
We validated the inferred $M_c$ through comparison with magnitude distributions in small areas over periods with constant station coverage and observed no significant bias.

\section*{Supplementary figures}
\FloatBarrier
\renewcommand\thefigure{S\arabic{figure}}
\renewcommand\thetable{S\arabic{table}}
\setcounter{figure}{0}
\setcounter{table}{0}

\begin{figure*}[ht!]
\centering
\includegraphics[width=\textwidth]{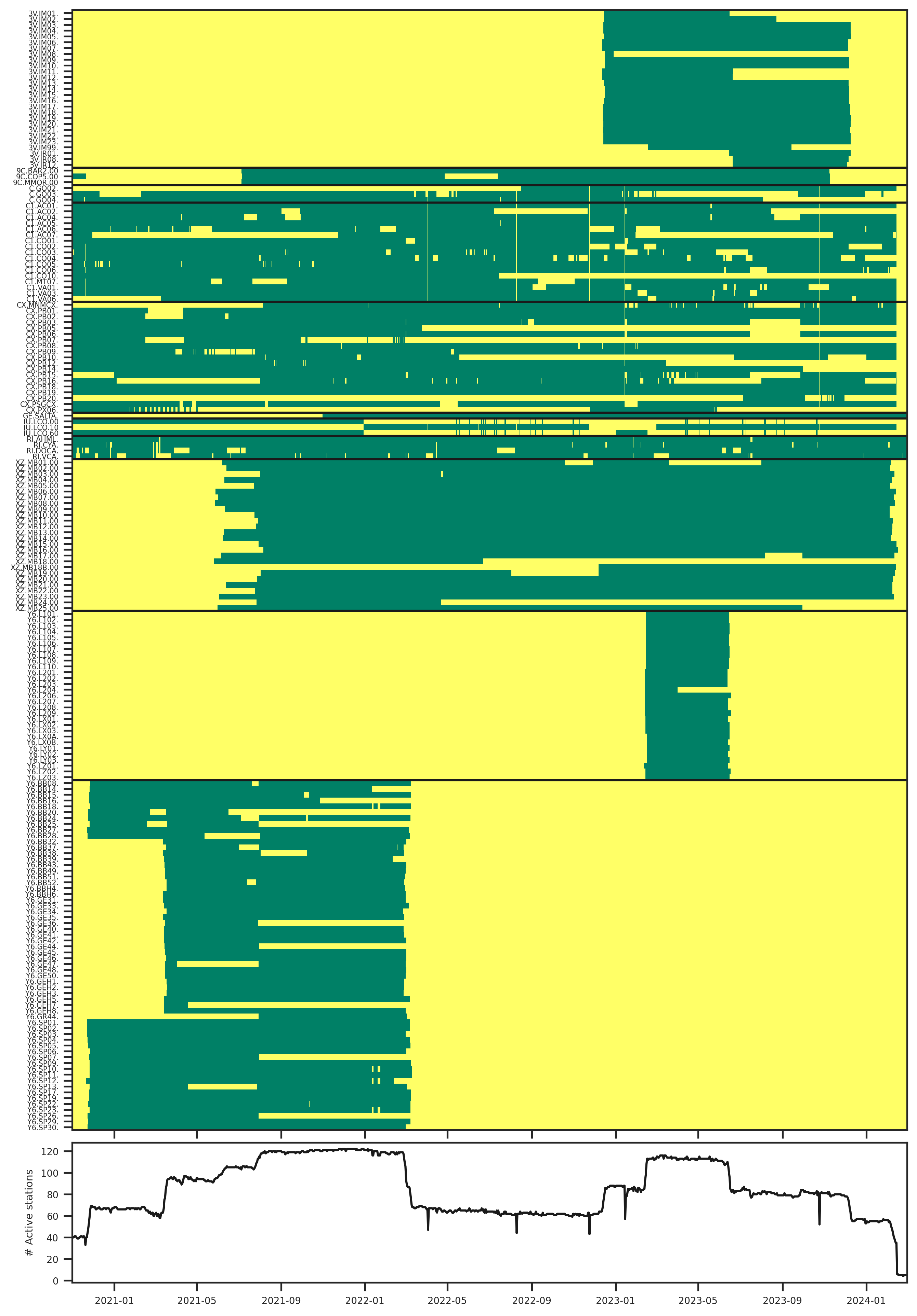}
\caption{Data availability for all stations. Green sections indicate available data, yellow sections indicate no data. Stations have been sorted alphabetically by network. Networks are separated by horizontal black lines. We split the Y6 network between the large-scale deployment and the later deployment during the active seismic cruise. The plot at the bottom shows the number of daily active stations over time. Data availability for the stations used for the longer-term catalog is shown in Figure~\ref{fig:data_availability_diego}.}
\label{fig:data_availability}
\end{figure*}

\begin{figure*}[ht!]
\centering
\includegraphics[width=\textwidth]{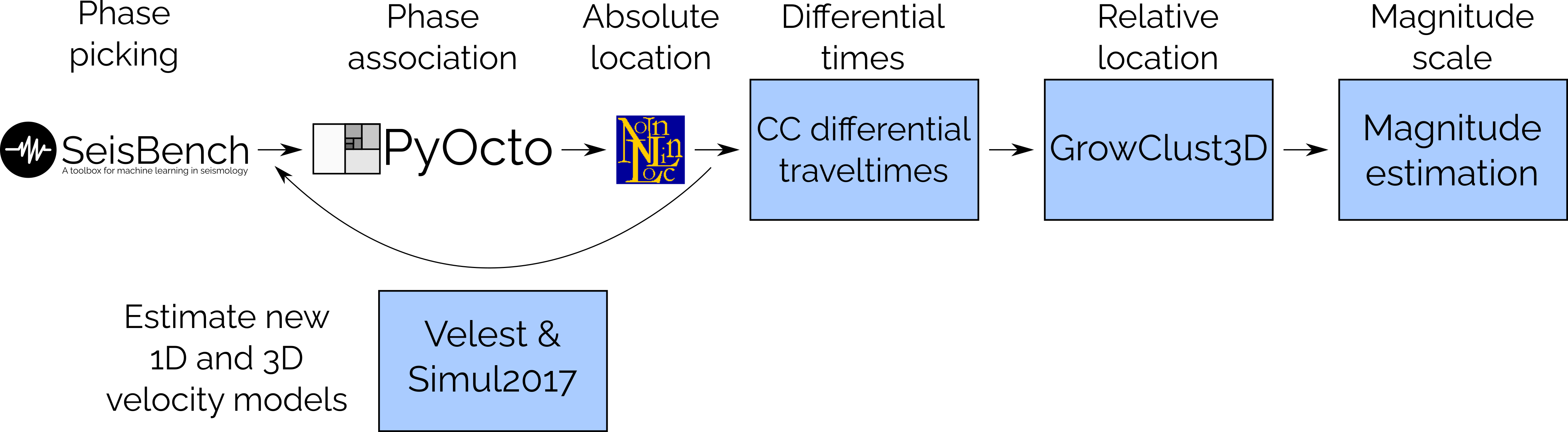}
\caption{Overview of the catalog generation workflow. The estimation of station residuals is not visualised.}
\label{fig:workflow}
\end{figure*}

\begin{figure*}[ht!]
\centering
\includegraphics[width=0.6\textwidth]{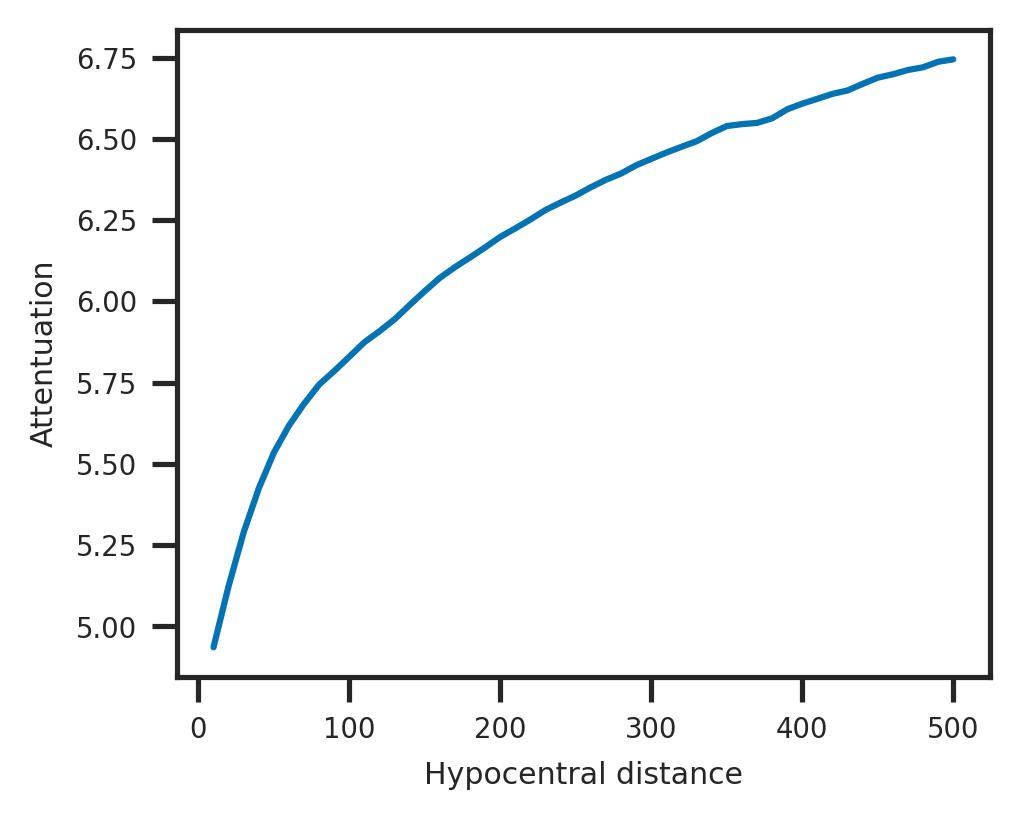}
\caption{Attenuation curve over hypocentral distance. The attenuation is denoted in log10 units and additive, i.e., to estimate magnitudes from a record at a fixed distance, the attenuation term needs to be added to the log10 of the Wood-Anderson response at that distance.}
\label{fig:attenuation_curve}
\end{figure*}

\begin{figure*}[ht!]
\centering
\includegraphics[width=\textwidth]{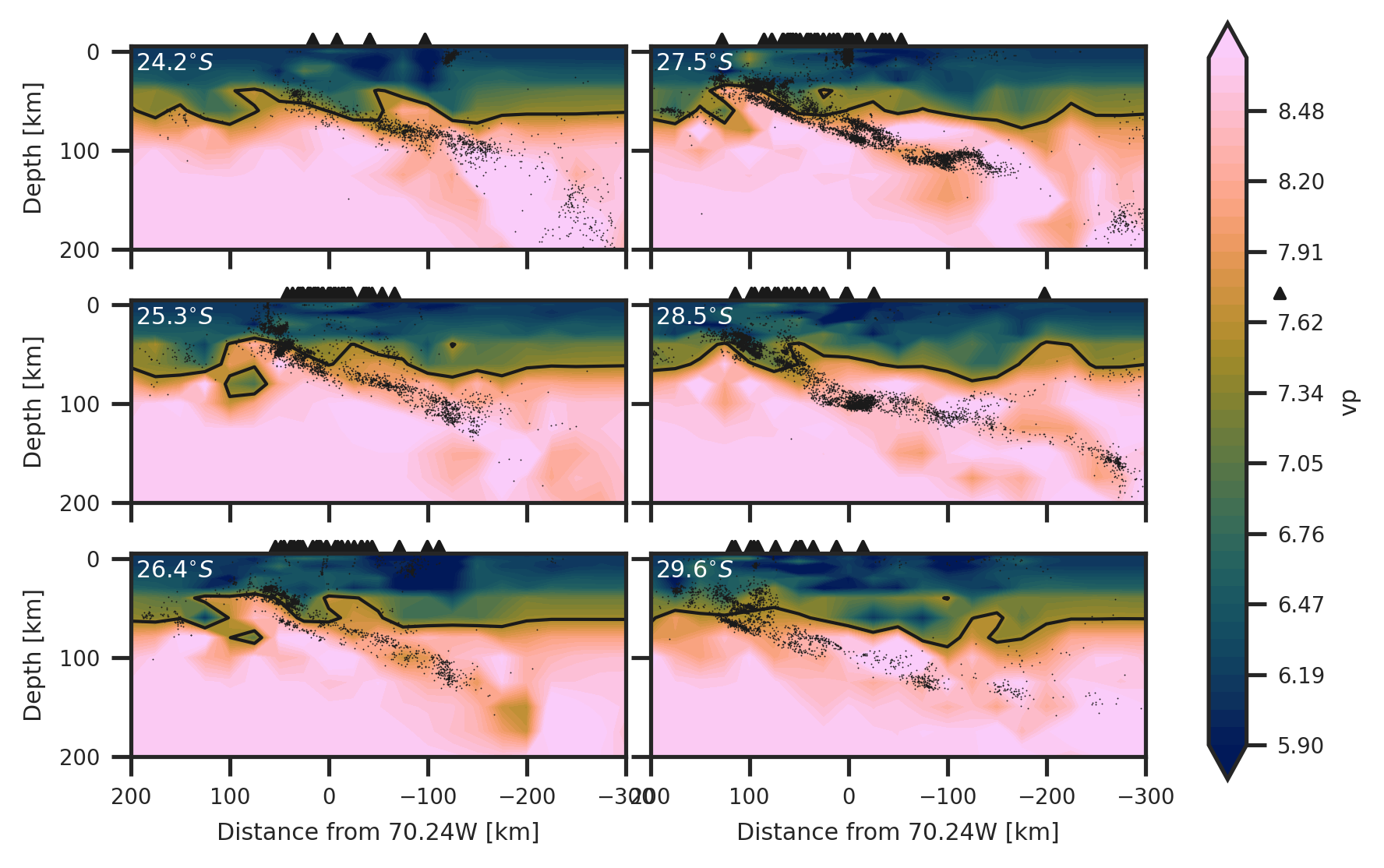}
\caption{3D P-wave velocity model inverted from the data and used for the location estimation. Black dots show a subset of the seismicity around the cross sections for orientation. The solid black line denotes 7.5~km/h isoline, a classic proxy for the Moho location. Black triangles on top show the station around the profiles.}
\label{fig:velocity_vp}
\end{figure*}

\begin{figure*}[ht!]
\centering
\includegraphics[width=\textwidth]{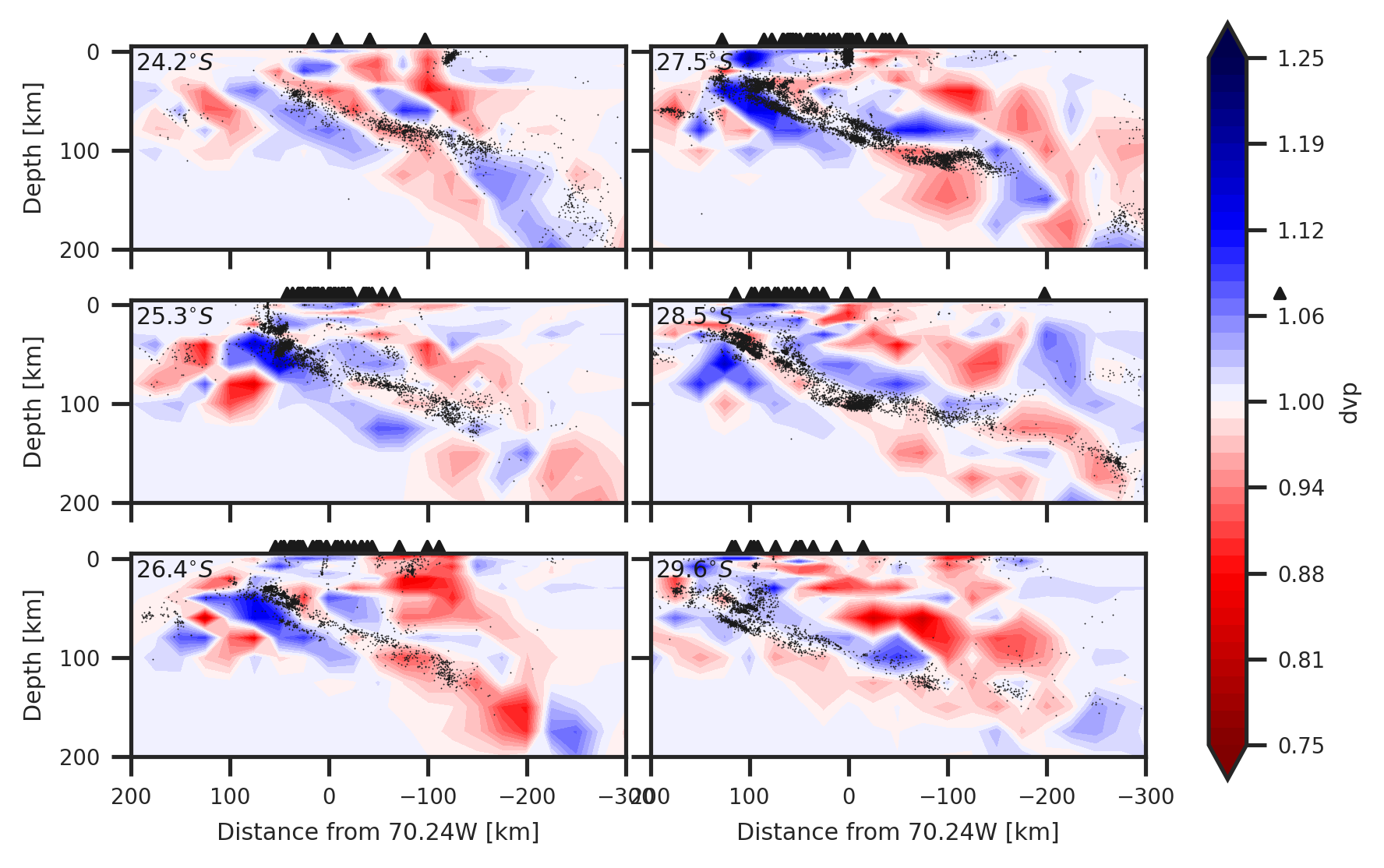}
\caption{Deviation of the P velocity from the average velocity at each depth. See Figure~\ref{fig:velocity_vp} for further details.}
\label{fig:velocity_dvp}
\end{figure*}

\begin{figure*}[ht!]
\centering
\includegraphics[width=\textwidth]{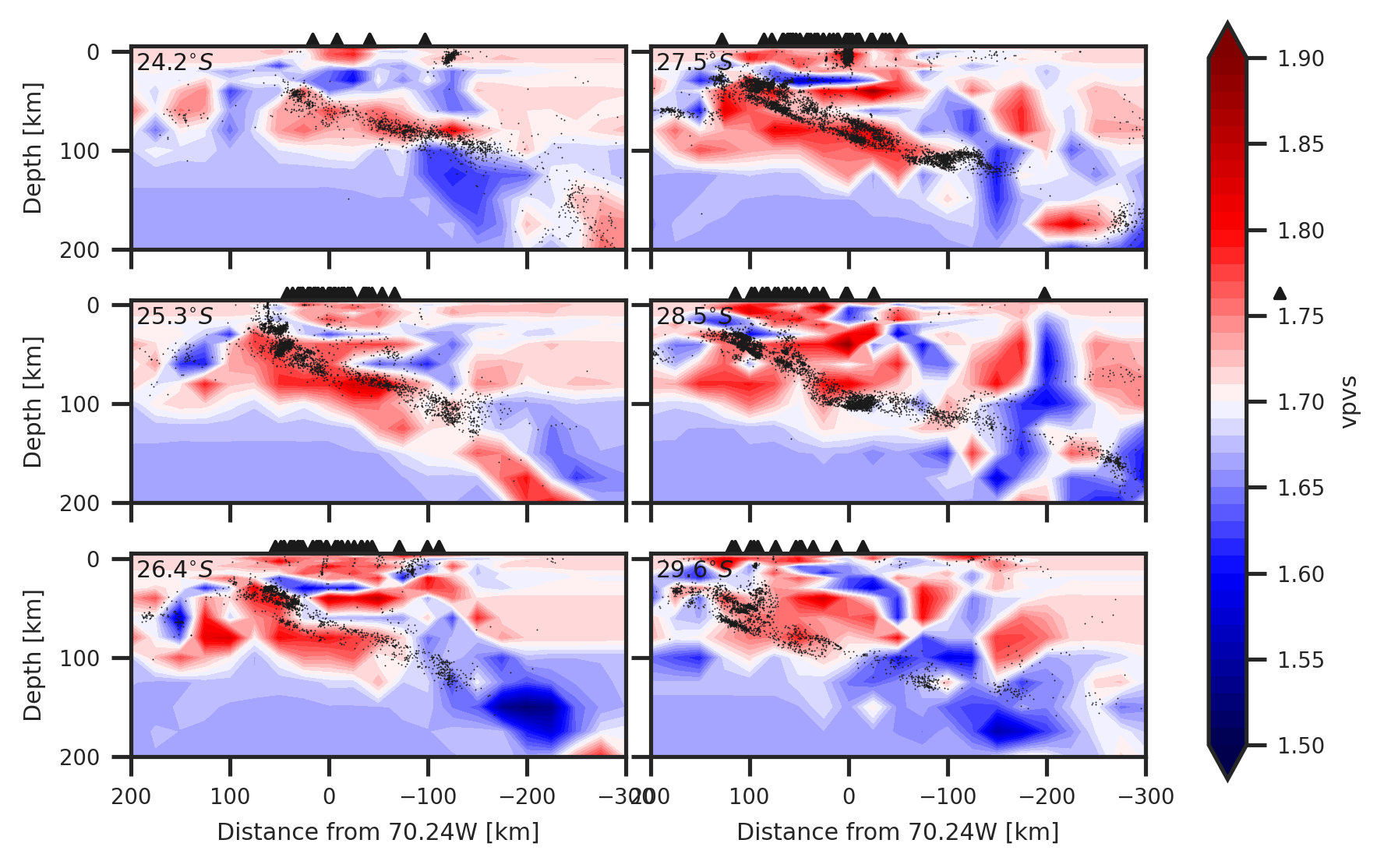}
\caption{Vp/Vs ratio for the 3D velocity model. See Figure~\ref{fig:velocity_vp} for further details.}
\label{fig:velocity_vpvs}
\end{figure*}

\begin{figure*}[ht!]
\centering
\includegraphics[width=1\textwidth]{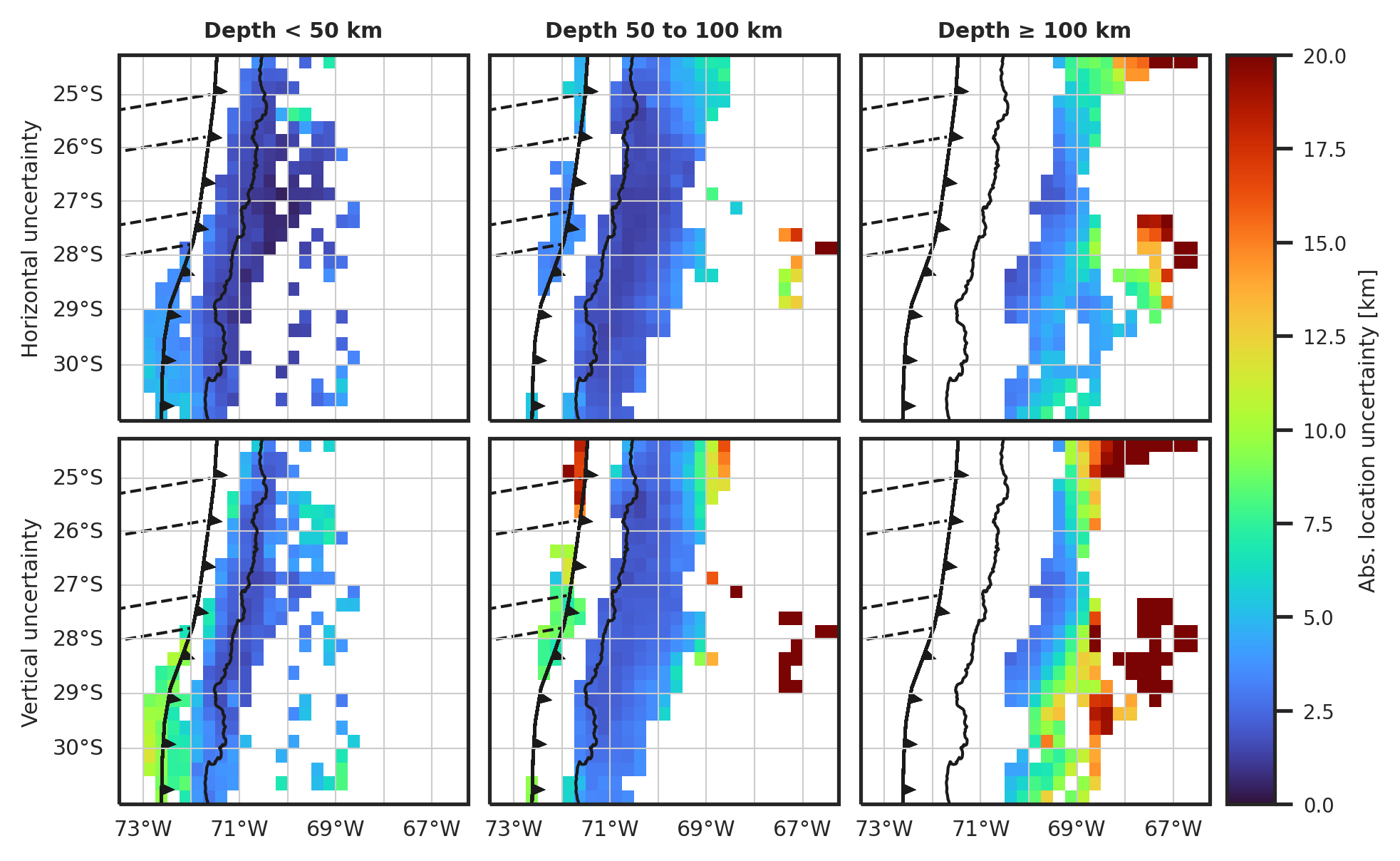}
\caption{Absolute horizontal (top) and vertical (bottom) location uncertainties from NonLinLoc (68\% confidence intervals). We report median uncertainties per 0.25\degree $\times$ 0.25\degree grid cells within three depth levels. Cells with less than 20 events are shown in white. Note that relative location errors after relocation with GrowClust will be substantially lower than the reported absolute uncertainties. Trench and coastline are plotted for orientation.}
\label{fig:nlloc_uncertainties}
\end{figure*}

\begin{figure*}[ht!]
\centering
\includegraphics[width=\textwidth]{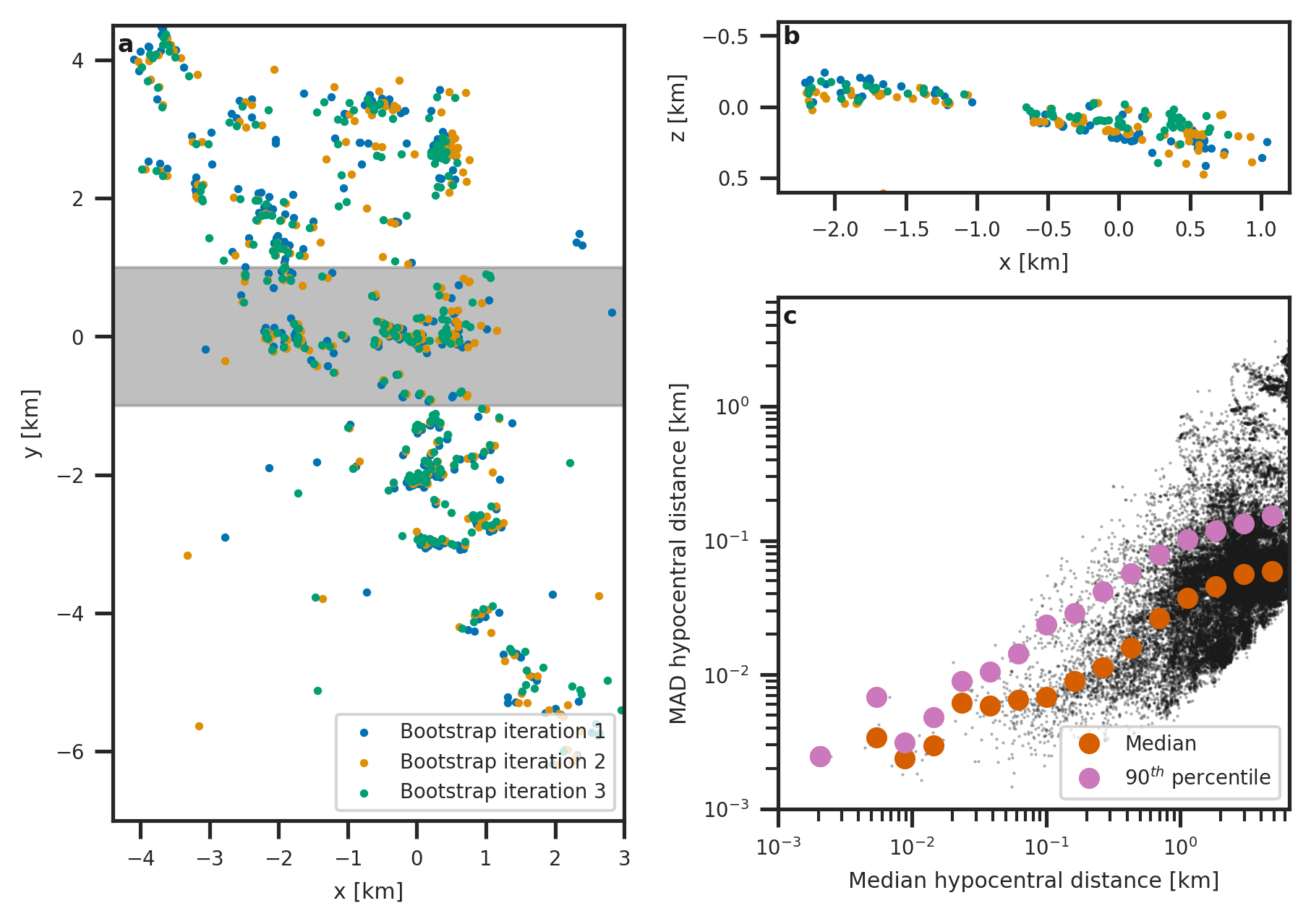}
\caption{Relocation bootstrap analysis with GrowClust3D. The analysis was conducted on all events located between 72\degree W and 71\degree W and between 28\degree S and 26.9\degree S. We show the initiation cluster of the 2023 SSE sequence around 71.3\degree W 27.3\degree S. All events have been transformed to a local coordinate system, the absolute location offset between the individual bootstrap iterations has been removed. \textbf{a} Epicenter distribution in the first 3 bootstrap iterations. \textbf{b} Cross section through the cluster. The events shown are highlighted in grey in subplot a. \textbf{c} Statistical results from 1000 bootstrap iterations. Each black dot represents one event pair with their median hypocentral distance and the median absolute deviation of the distance as an uncertainty measure. Red and purple dots show the moving median and $90^{th}$ percentile of the uncertainties.}
\label{fig:relocation_bootstrap}
\end{figure*}

\begin{figure*}[ht!]
\centering
\includegraphics[width=0.6\textwidth]{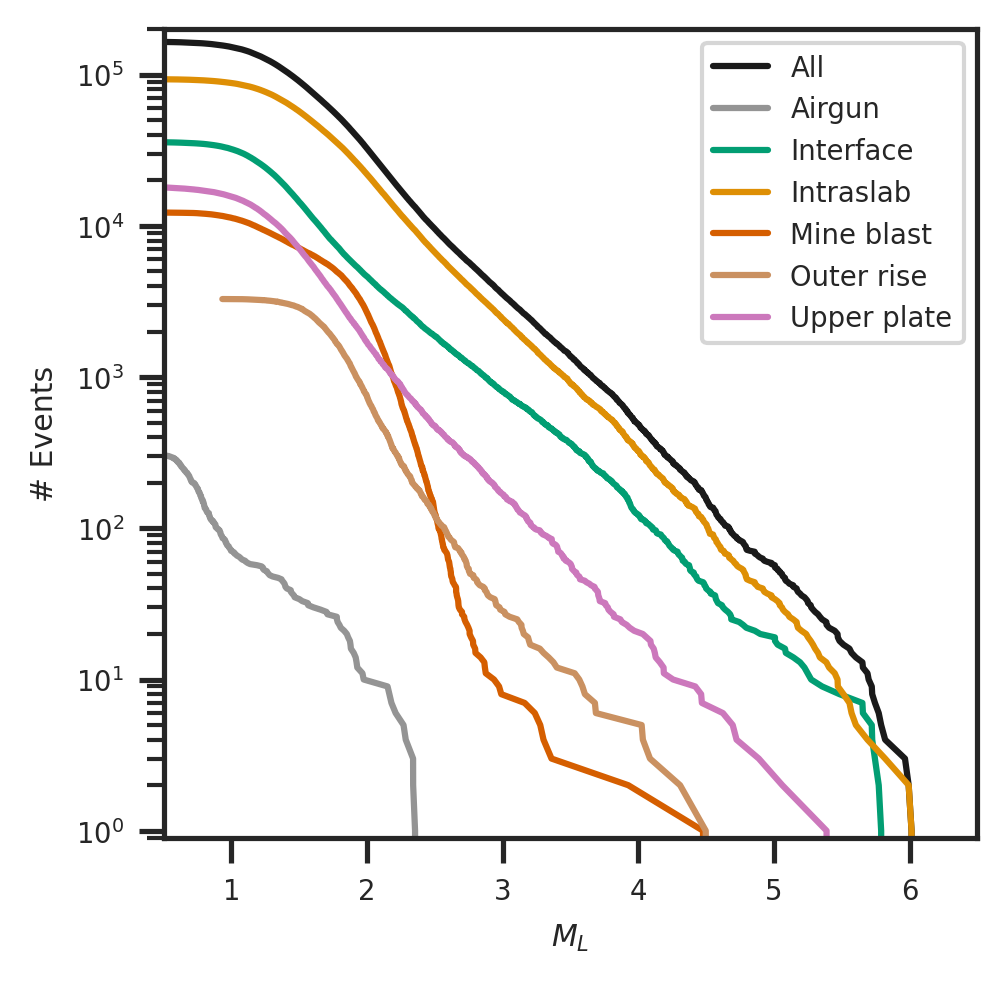}
\caption{Inverse cumulative distribution of the magnitudes by event category. We record $<10$ mine blasting events with an apparent magnitude above $3.0$. These magnitudes are likely overestimations.}
\label{fig:magnitude_histogram}
\end{figure*}

\begin{figure*}[ht!]
\centering
\includegraphics[width=\textwidth]{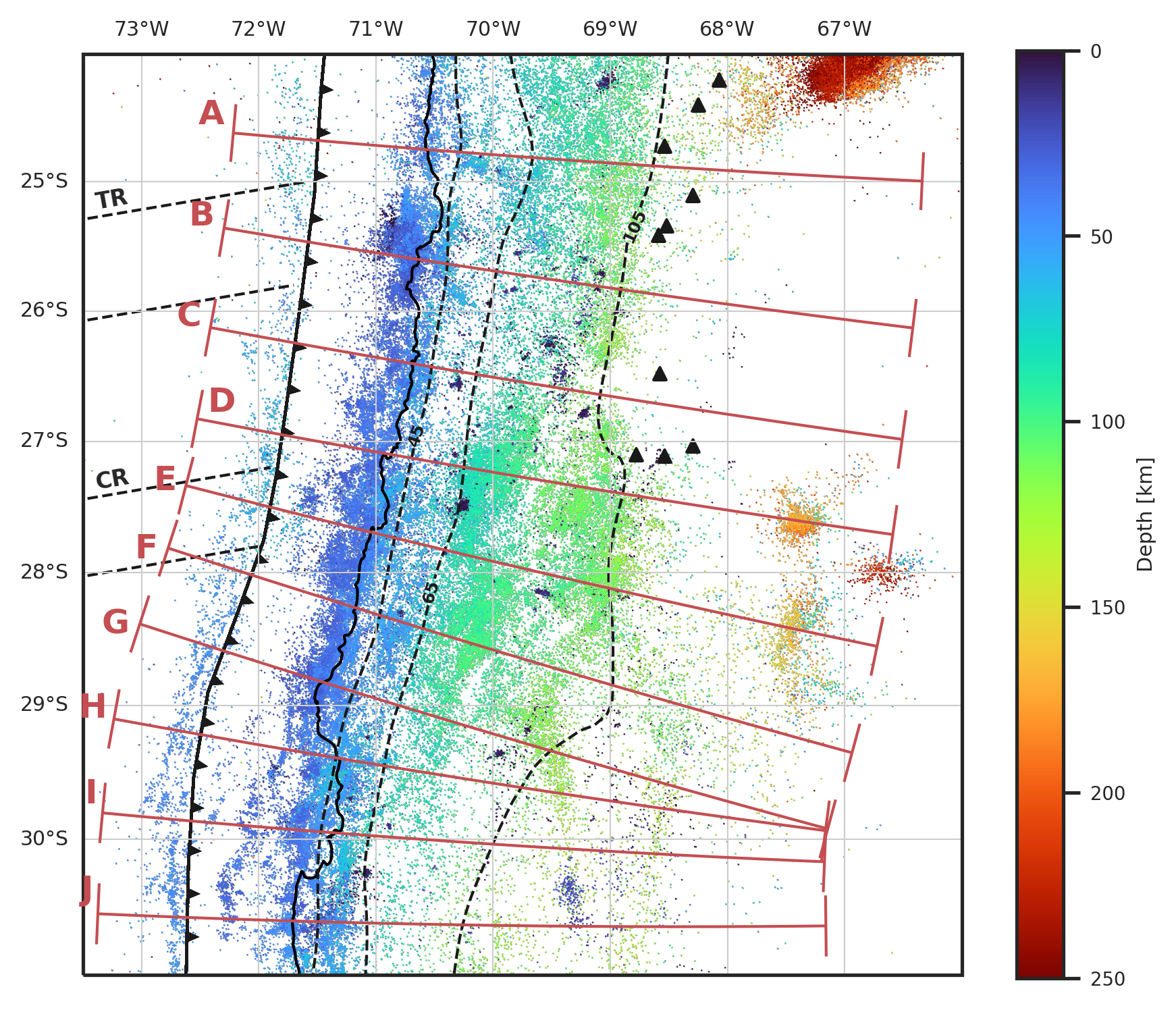}
\caption{Overview of the seismicity as in Figure~\ref{fig:overview}. The indicated cross sections are shown in Figure~\ref{fig:cross_sections_fine}.}
\label{fig:events_overview_fine}
\end{figure*}

\begin{figure*}[ht!]
\centering
\includegraphics[width=\textwidth]{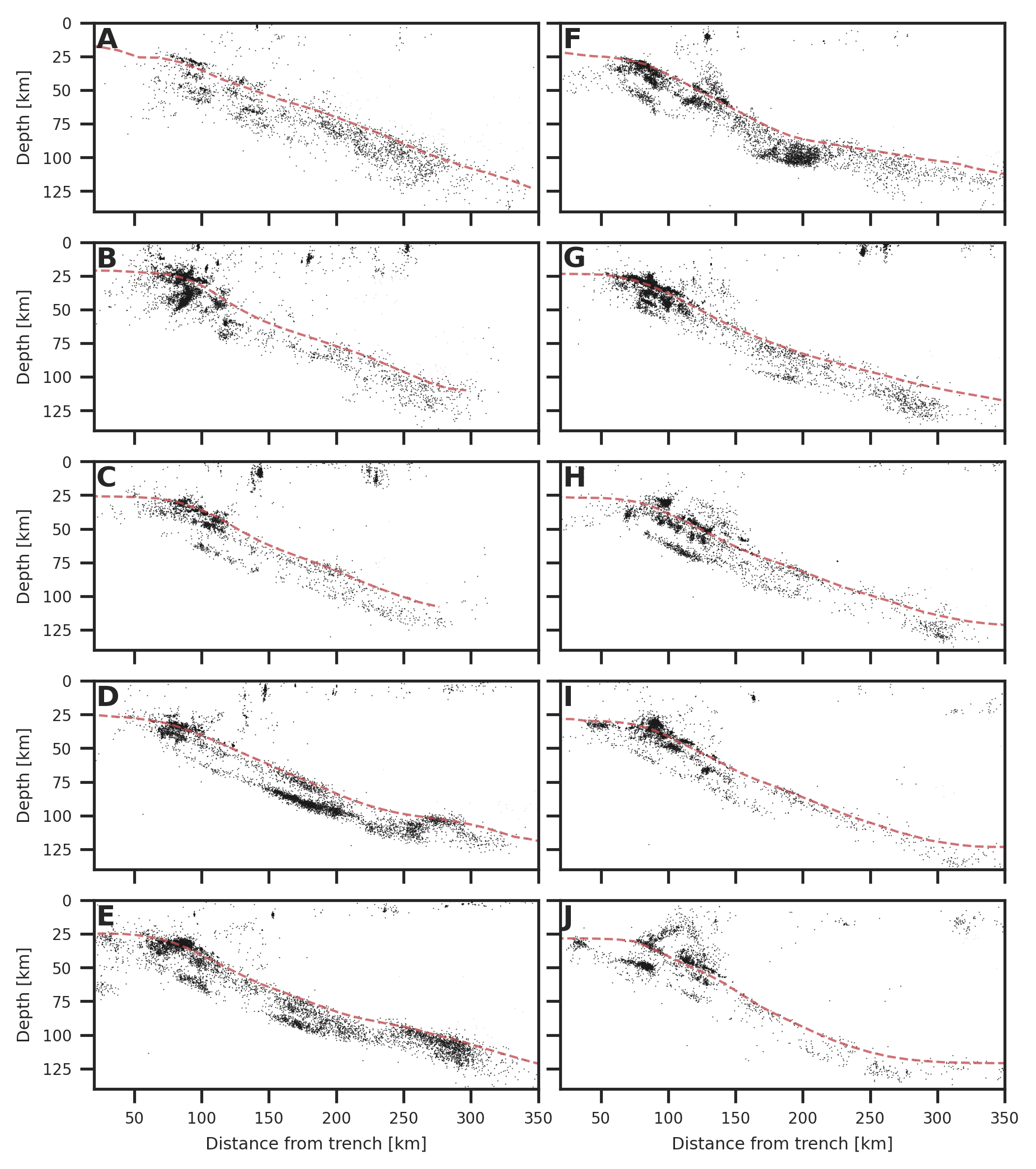}
\caption{Trench-perpendicular cross sections of the seismicity. The slab model (dashed red line) has been inferred from the seismicity. The location of the cross sections in shown in Figure~\ref{fig:events_overview_fine}. All cross-sections have a total width of 30~km, i.e., 15~km to each side of the center.}
\label{fig:cross_sections_fine}
\end{figure*}

\begin{figure*}[ht!]
\centering
\includegraphics[width=\textwidth]{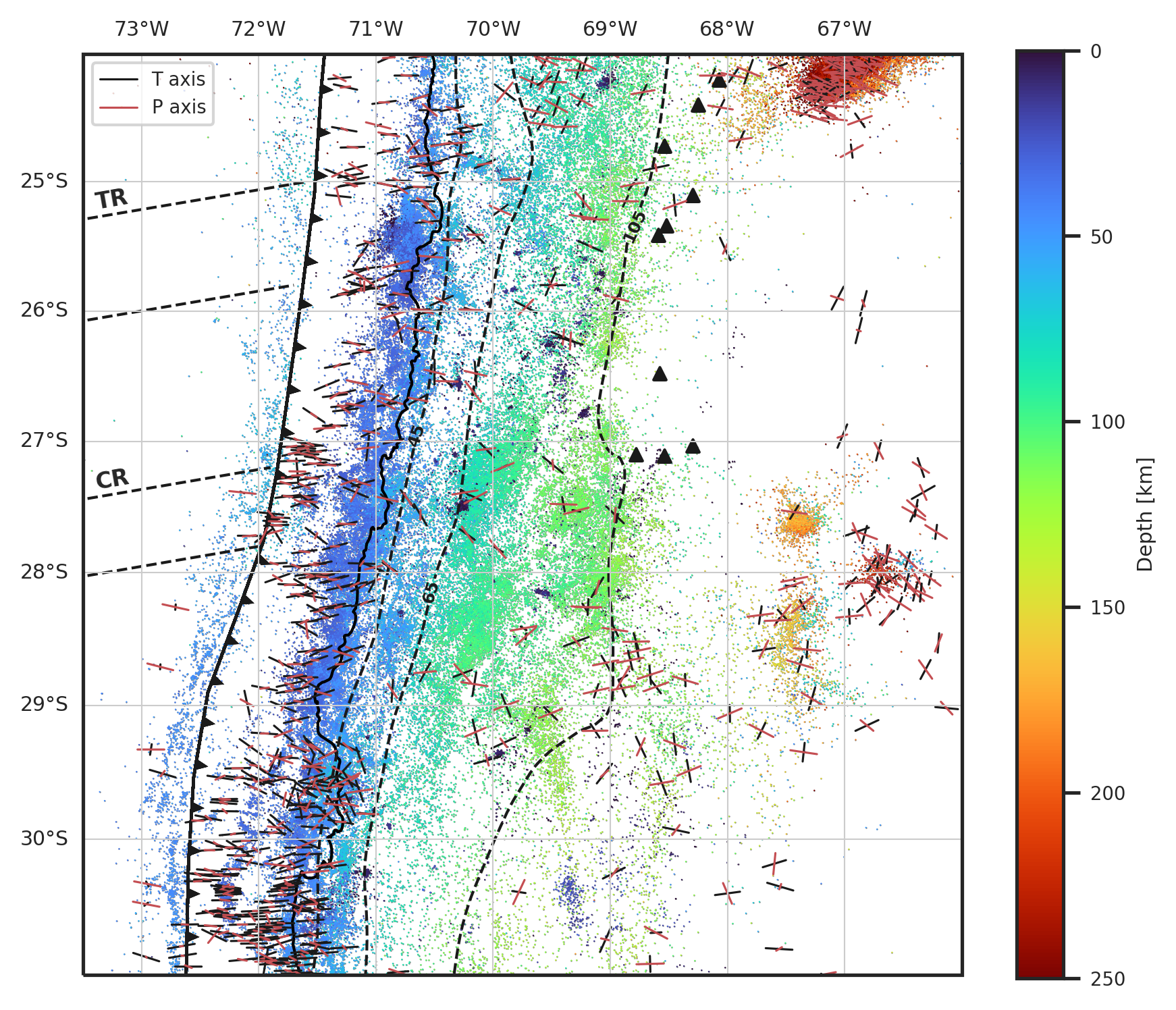}
\caption{Same as Figure~\ref{fig:overview} (left) but with moment tensors from GCMT \cite{ekstromGlobalCMTProject2012} plotted on top of it. Moment tensors are visualised by their P (red) and T (black) axis. We use all moment tensors available in GCMT, including the ones before or after our study period. We plot the events at the centroid location provided by GCMT, determined based on teleseismic waves. This leads to an apparent offset between GCMT events and structures mapped with our local catalog.}
\label{fig:events_overview_mt}
\end{figure*}

\begin{figure*}[ht!]
\centering
\includegraphics[width=\textwidth]{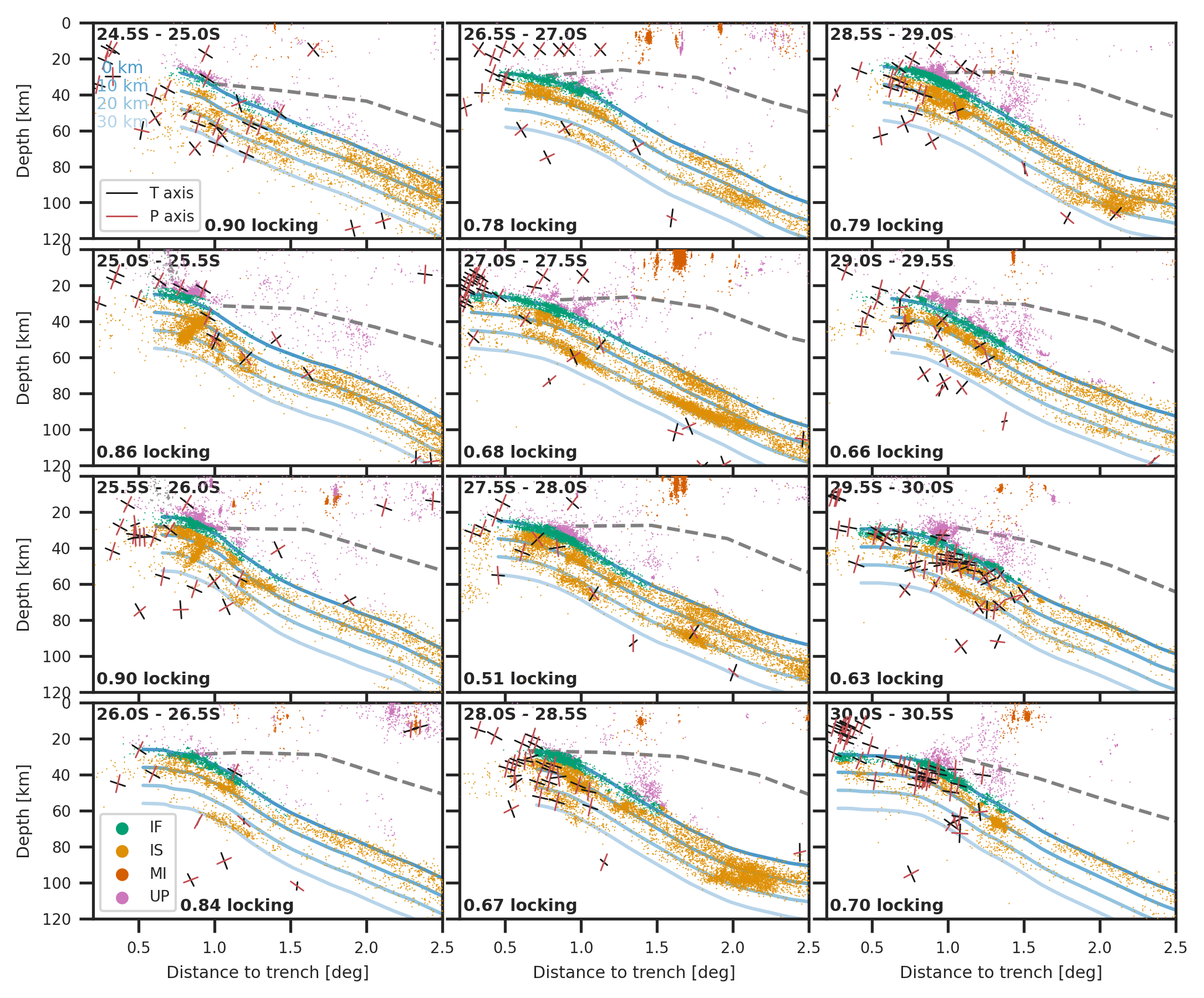}
\caption{Same as Figure~\ref{fig:cross_section_ortho} but with moment tensors from GCMT \cite{ekstromGlobalCMTProject2012} plotted on top of it. Moment tensors are visualised by their P (red) and T (black) axis. We use all moment tensors available in GCMT, including the ones before or after our study period. We plot the events at the centroid location provided by GCMT, determined based on teleseismic waves. This leads to an apparent offset between GCMT events and structures mapped with our local catalog.}
\label{fig:cross_section_ortho_mt}
\end{figure*}

\begin{figure*}[ht!]
\centering
\includegraphics[width=\textwidth]{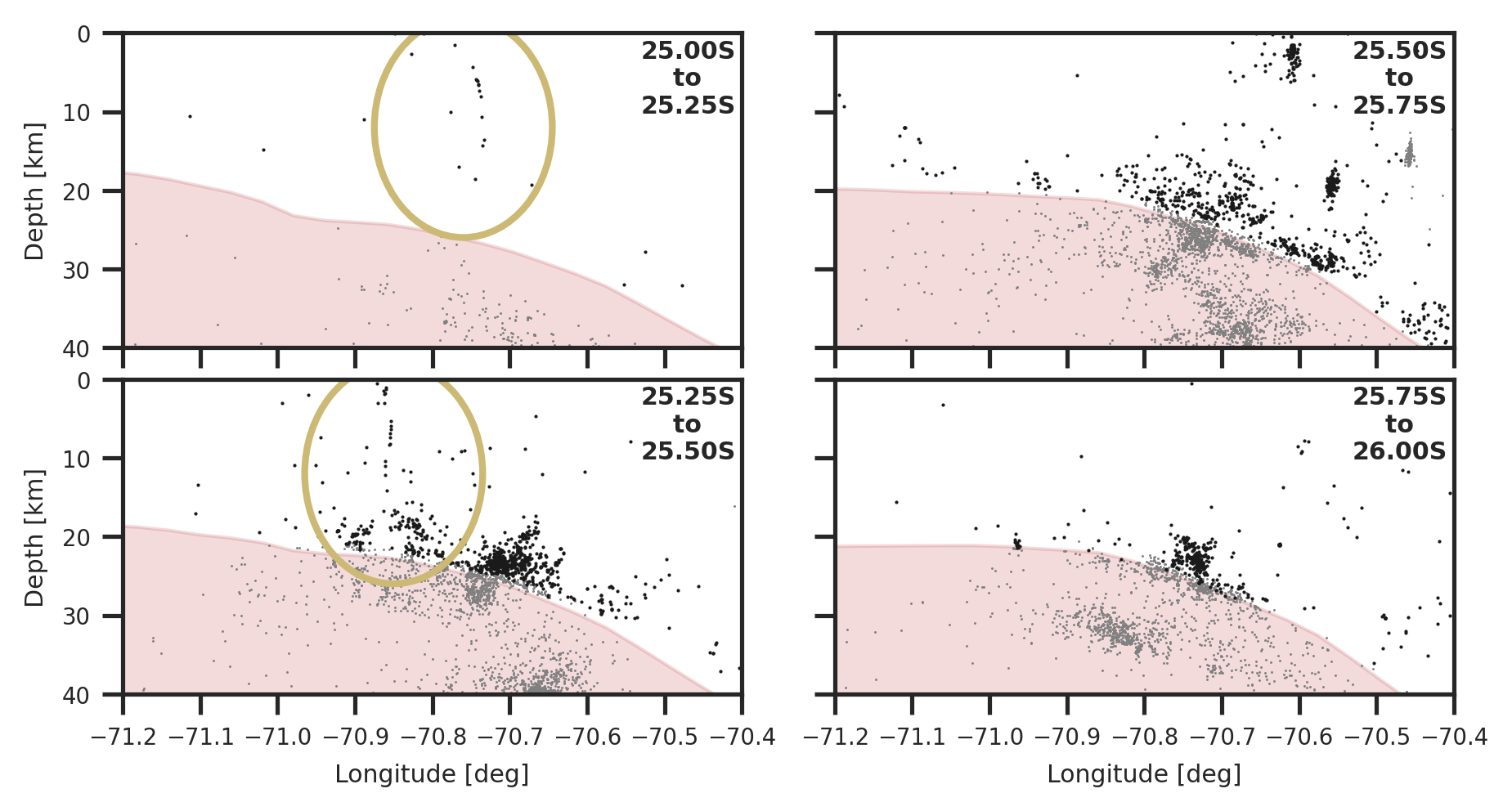}
\caption{Cross-section of the crustal seismicity in the area of Taltal. Yellow ellipses highlight a potential splay structure. The approximate extent of the subducting plate is indicated in red.}
\label{fig:cross_section_crustal}
\end{figure*}

\begin{figure*}[ht!]
\centering
\includegraphics[width=\textwidth]{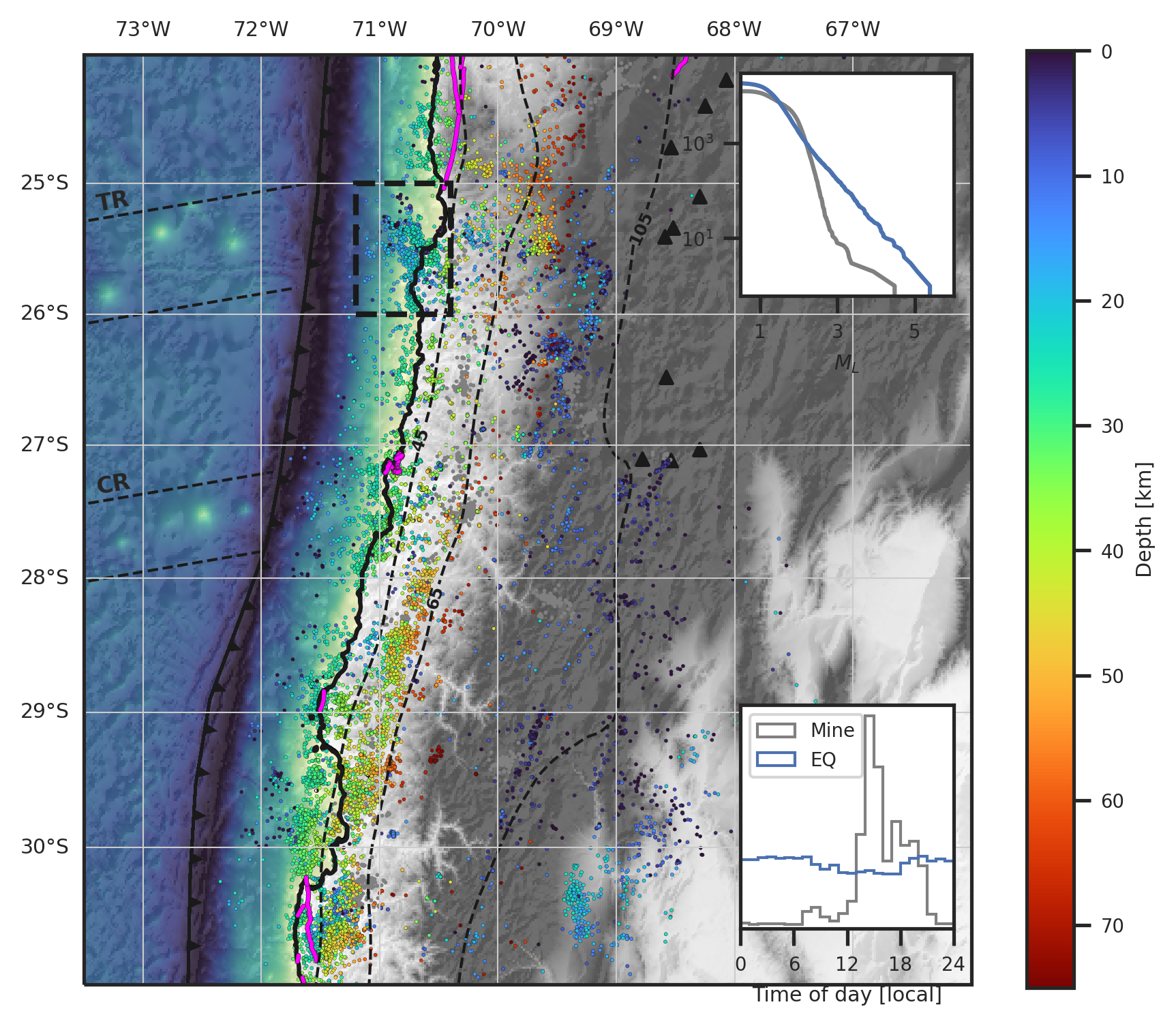}
\caption{Same as Figure~\ref{fig:upper_plate} but showing the bathymetry and topographic and omitting the plate locking.}
\label{fig:upper_plate_topography}
\end{figure*}

\begin{figure*}[ht!]
\centering
\includegraphics[width=\textwidth]{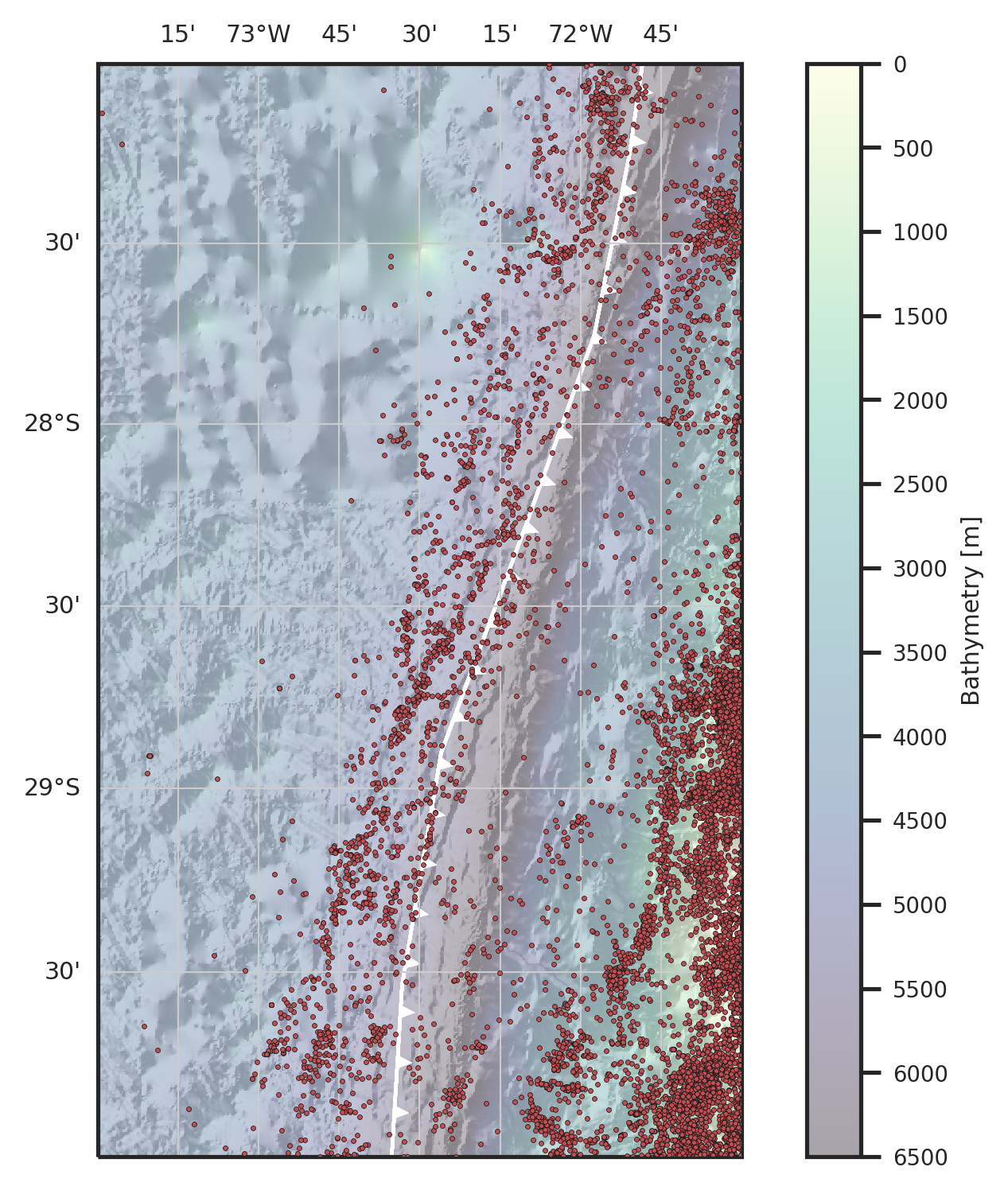}
\caption{Events and bathymetry offshore in the Southern part of the study region. Earthquakes are denoted by red dots, the bathymetry in yellow to blue colors. For the bathymetry, we use a light direction of 120\degree ~to highlight the seafloor fault traces.}
\label{fig:outer_rise}
\end{figure*}

\begin{figure*}[ht!]
\centering
\includegraphics[width=\textwidth]{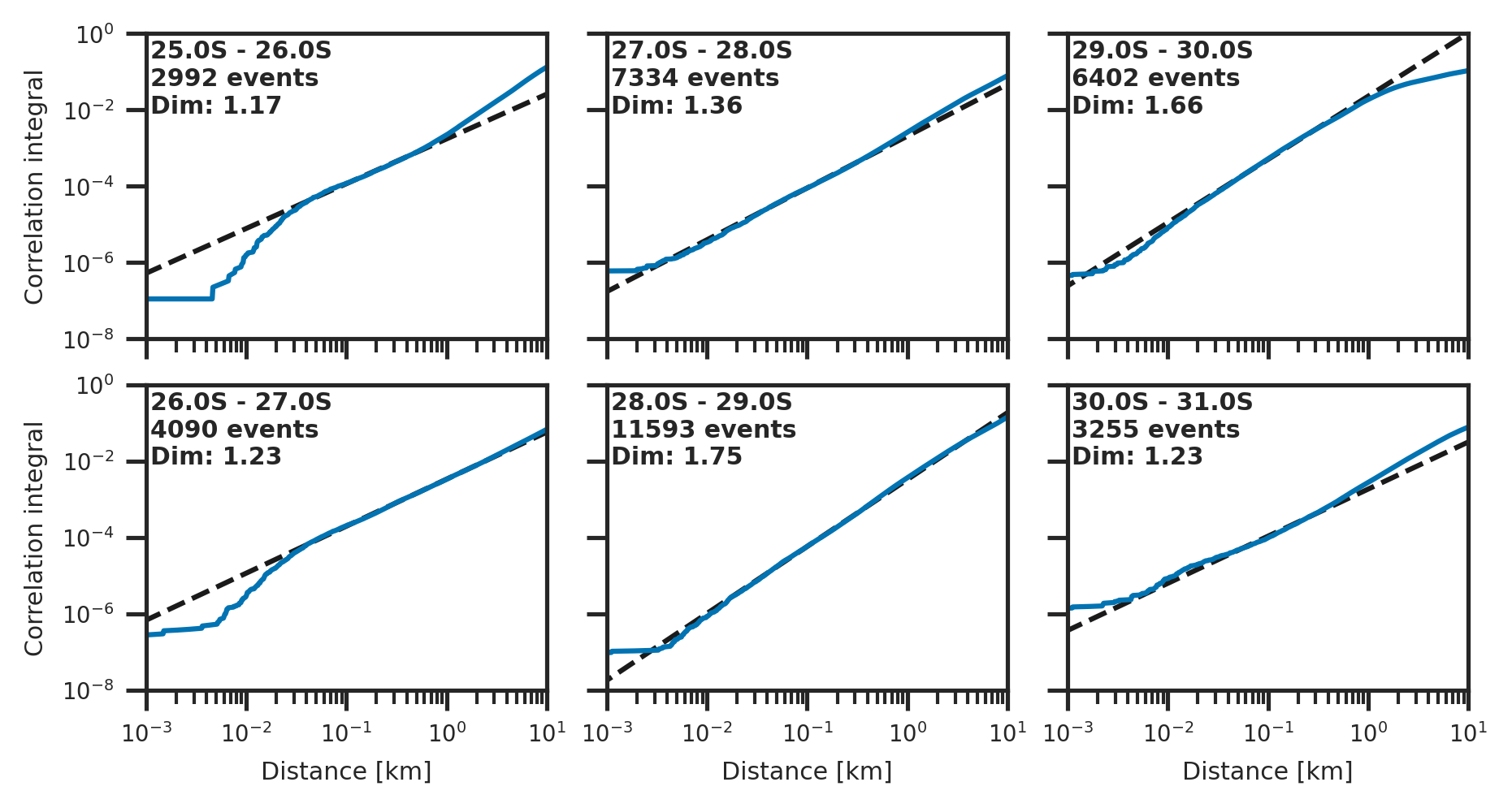}
\caption{Correlation integral for interface events grouped by latitude. In the upper left corner, we provide the latitude range, the number of events, and the estimated fractal dimension. We estimate the fractal dimension using a log-log fit between 30~m and 500~m, as we observed consistent power law behaviour in this range. We calculate the correlation integral from the hypocentral positions, i.e., including the event depth.}
\label{fig:interface_fractal_3d}
\end{figure*}

\begin{figure*}[ht!]
\centering
\includegraphics[width=\textwidth]{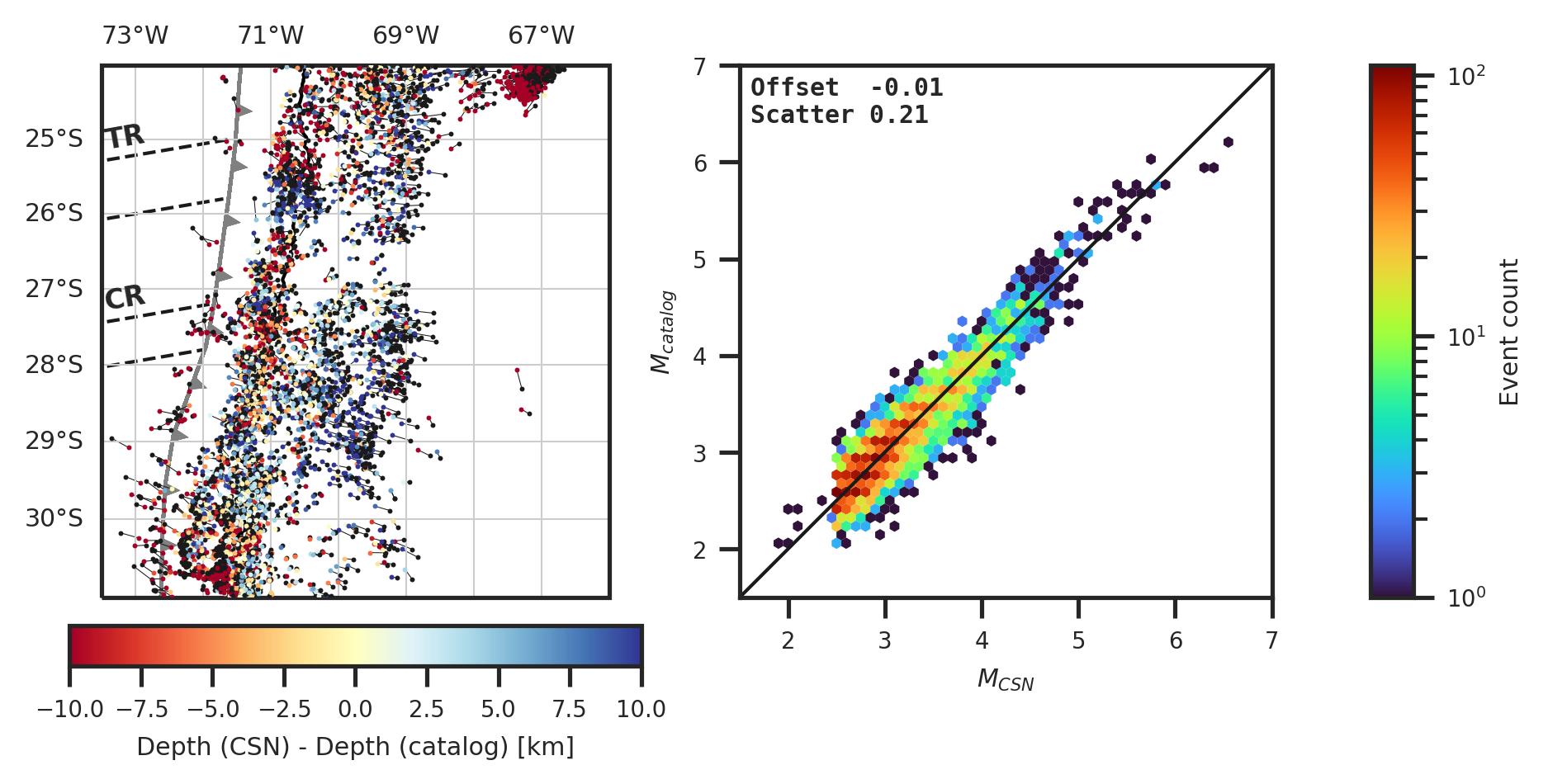}
\caption{Comparison between the detected events and the earthquake catalog from the Chilean Seismic Network (CSN). Events are matched based on origin times, with a maximum time difference of 5~s. In the map, our events are denoted in black, CSN events in color, with the color encoding the depth difference between the events. For red dots, the CSN location is shallower than our location, for blue dots the CSN location is deeper. Matching events are connected by black lines. The magnitude comparison shows a density plot. Values in the corner indicate the mean difference (offset) and the standard deviation (scatter).}
\label{fig:csn_comparison}
\end{figure*}

\begin{figure*}[ht!]
\centering
\includegraphics[width=\textwidth]{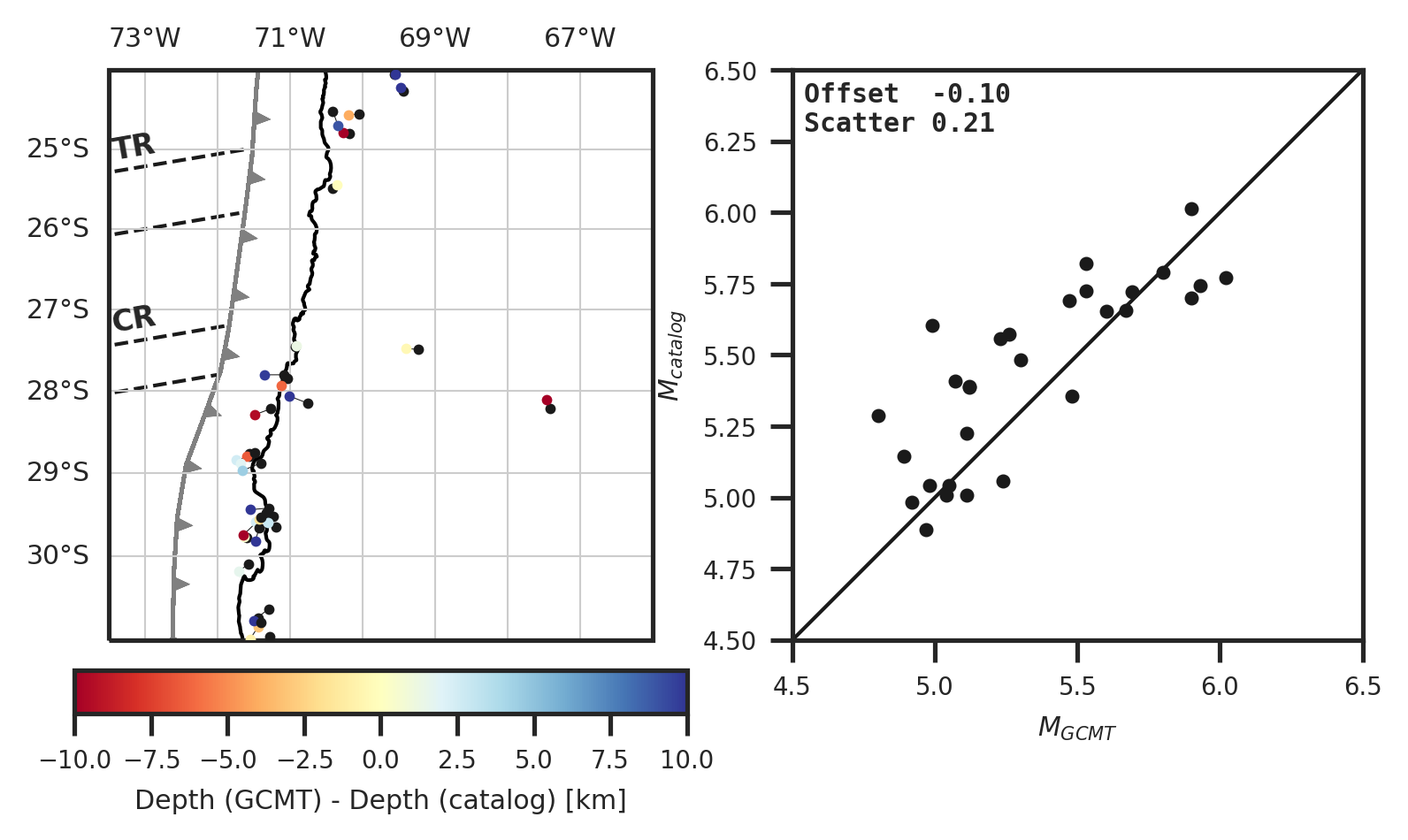}
\caption{Comparison between the detected events and the GCMT catalog. Events are matched based on origin times, with a maximum time difference of 5~s. In the map, our events are denoted in black, GCMT events in color, with the color encoding the depth difference between the events. For red dots, the GCMT location is shallower than our location, for blue dots the GCMT location is deeper. Matching events are connected by black lines. The magnitude comparison shows a scatter plot. Values in the corner indicate the mean difference (offset) and the standard deviation (scatter).}
\label{fig:cmt_comparison}
\end{figure*}

\begin{figure*}[ht!]
\centering
\includegraphics[width=\textwidth]{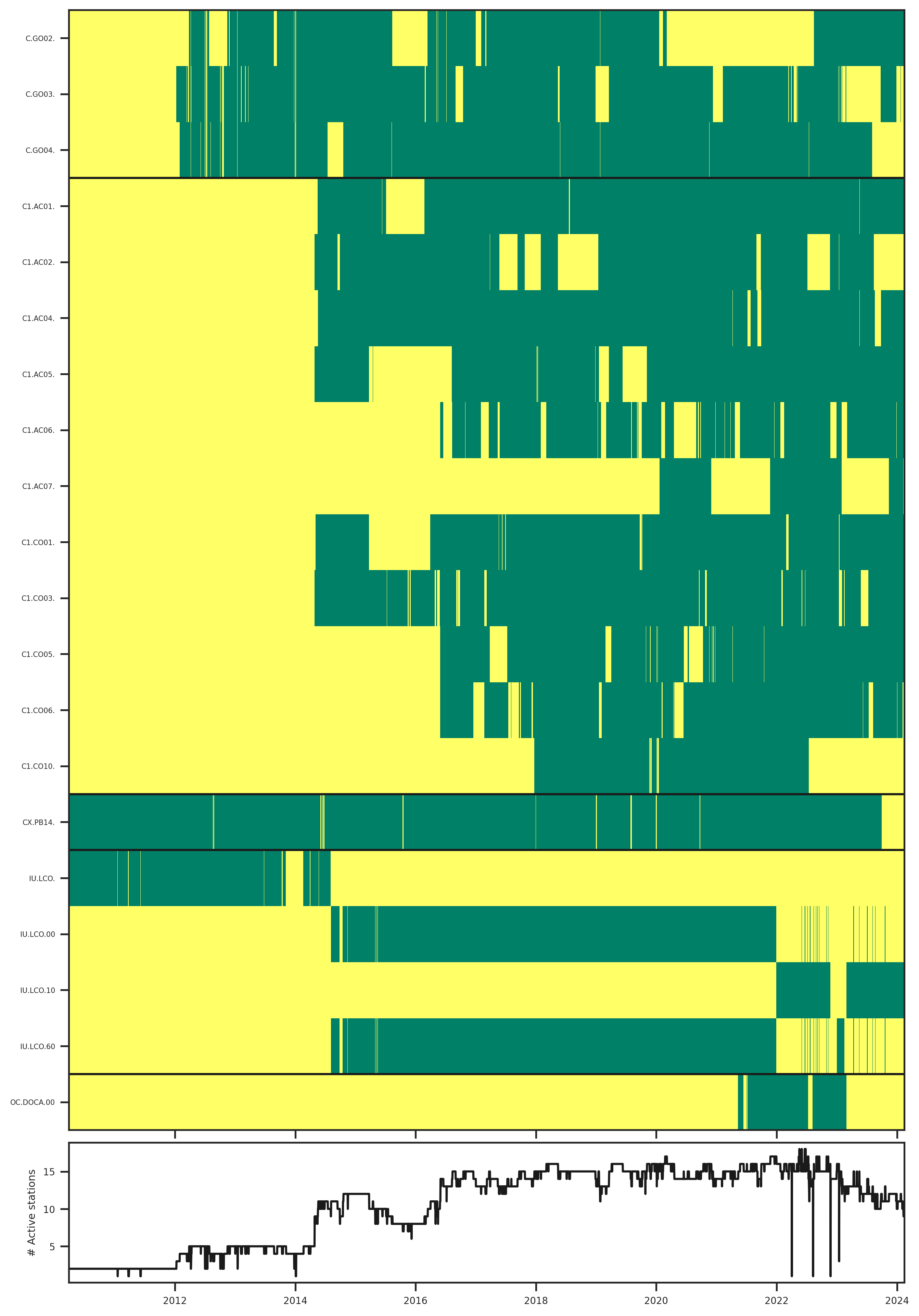}
\caption{Data availability for the stations from the permanent catalog. See Figure~\ref{fig:data_availability} for further details.}
\label{fig:data_availability_diego}
\end{figure*}

\end{document}